\documentclass[aps,showpacs,showkeys,nofootinbib,twocolumn]{revtex4}
\usepackage{epsfig,amssymb,bm,graphics,color}
\newcommand*{\fs}[1]{#1\!\!\!/}
\newcommand*{\fsk}{k\!\!\!/}
\newcommand*{\ee}{e^+e^-}

\begin{document} {\normalsize }
\title{
Polarization of recoil photon in  non-linear Compton process  
}
 \author{A.~I.~Titov}
 \affiliation{  
   Bogoliubov Laboratory of Theoretical Physics,
   Joint Institute for Nuclear Research, Dubna 141980, Russia
\footnote{email:atitov@theor.jinr.ru}}
 \begin{abstract}
   The polarization of recoil photon ($\gamma'$) in the non-linear
   Compton process $e + \vec L \to \vec \gamma' +e'$ in the interaction           
   of a relativistic electron with a linearly polarized laser beam ($\vec L$)
   is studied
   within the Furry picture in the lowest-order, tree-level S matrix
   element.
   In particular, we consider the asymmetry of differential
   cross sections ${\cal A}$ for two independent
   axes describing the Compton process
   equal to the intrinsic spin variable ${\xi}^f_3$, that determines                
   the polarization properties of $\gamma'$.
   The sign and absolute value of the asymmetry determine
   the direction and degree of $\gamma'$ polarization.  
   We have analyzed the process  in a wide range of laser intensity
   that covers existing and future experiments.
   Our results provide additional knowledge for studying non-linear
   multi-photon effects in quantum electrodynamics and can be used
   in planning experiments at envisaged laser facilities.
 \end{abstract}
\maketitle

\section{Introduction}
The dynamics of the interaction of an ultra-relativistic electron beam
with a super-intense laser attracts much attention from theory
and experiment in the hope of shedding light on open problems
of non-linear quantum electrodynamics.
An excellent analysis of theoretical
achievements and possible expectations from new facilities is given
in a recent review ~\cite{AM_Review}, see also~\cite{ADPiazza}.
Important parts of these studies are the non-linear Compton (nlCo) and
non-linear Breit-Wheeler (nlBW) processes, where a probe particle,
electron or photon, respectively, interacts with a high-intensity
background electromagnetic field (laser).
Both processes were thoroughly researched theoretically
in the past and then re-considered and improved~
\cite{RitusGroup,Ritus-79,Serbo2004,Seipt:2020diz,TitovPRA2013,DiPiazza:2020wxp,Ilderton:2020dhs,Piazza2018,Heinzl2020,TitovPEPAN,Granz:2019sxb,Kampfer2023}.
In most cases, the field of a high-intensity optical laser
is considered as a background field. For example, the famous
SLAC experiment E-144~\cite{E-144},
envisaged European LUXE~\cite{LUXE_exp} and
E-320 at FACET~II/SLAC~\cite{E_320,E_320add,E_320add23} projects.
For X-ray photon beam experiments (XFEL), see~\cite{XFEL}.
In nlBW, the formation of an electron-positron pair using
an optical laser pulse of several eV requires an external
photon with a frequency of tens of GeV.
This may be a recoil photon ($\gamma'$) arising from the
interaction of a relativistic electron with an optical laser beam
in a non-linear multiphoton Compton process.
Consideration of this process with an emphasis on
$\gamma'$-polarization is the main goal of this work.

We use the following notation.
The electromagnetic field intensity is described by the dimensionless
parameter
{$\xi=|e|{\cal E}/(m\omega)$, where
$\cal{E}$ and $\omega$ are the central electric field strength
and the central laser frequency, respectively;}
$-|e|$ and $m$ stand for electron charge and mass, respectively.
The four-vectors
$p=(E_e,\vec p)$ and $p'=(E'_e,\vec p')$ are the four-momenta
of incoming and outgoing bare electrons, respectively;
$q=(q_0,\vec{q})$ and $q'=(q'_0,\vec q')$ are the four-momenta for dressed
electrons~\cite{LL4}, respectively, with e.g., 
$q=p-k(\xi^2m^2/4k\cdot p)$,
where $k=(\omega,\mathbf k)$
is the four-momentum of the beam photon with $\mathbf{k}=\mathbf{z}\omega$. 

The four-momentum of the outgoing photon with frequency $\omega'$
is $k'=(\omega',\mathbf{k}')$ with
$\mathbf{k}'=\omega'(\mathbf{x}\sin\theta\cos\varphi
+\mathbf{y}\sin\theta\sin\varphi+\mathbf{z}\cos\theta)$, where
$\theta$ and $\varphi$ are the corresponding polar and azimuthal angles,
respectively.
We also use the quantum non-linearity parameter $\chi=\xi(k\cdot p)/m^2$.
It is assumed that the {laser} beam polarization
is directed along the $\mathbf{x}$ axis.
We use natural units with $c=\hbar=1$,
and $e^2/4\pi=\alpha\approx 1/137.036$.
{Following the LUXE experiment design,
we are limited to unpolarized electrons.}

In case of linearly polarized initial photons, the
yield of the $\ee$ pair in nlBW process depends on the kinematics (square of the
total energy in c.m s., $s$), $\xi$, and  mutual polarization of
the initial photons.
For example, at ultra-high field intensity with $\xi\gg1$
the asymmetry of the electron-positron yield
${\cal{A}}=(W_{\perp}-W_{\parallel})/(W_{\perp}+W_{\parallel})$,
when the polarization of the initial
photons are mutually perpendicular $W_{\perp}$
or parallel $W_{\parallel}$,
varies from 1/3 to 1/5~\cite{Ritus-79},
depending on $\kappa=\xi s/2m^2$.
At low field strength with $\xi\lesssim 1$, the asymmetry
$\cal{A}$ exhibits non-monotonic behavior, varying from zero
to unity depending on the initial kinematic conditions~\cite{TK2020}.
Therefore, the mutual polarization of two linearly polarized
initial photons is an important component of the nlBW process.
Since nlCo is a possible source of high energy photons, studying
the polarization of Compton recoil photons is useful. In addition,
it has independent significance as a source of additional
information about the dynamics of non-linear multiphoton processes.

In the nlCo process, the polarization density
matrix for the recoil photon $\rho^f$ is expressed through the
amplitude~\cite{Akhiezer_Berestetsky}
\begin{eqnarray}
  M=\sum\limits_{a} {e'_a}^*M(a)~,
\end{eqnarray}
where ${e'_a}$ is the polarization unit vector
for the photon in the final state as
\begin{eqnarray}
\rho^f_{ab}=\frac{M(a)M^*(b)}{\sum\limits_{a}|M(a)|^2}~.
\end{eqnarray}
The corresponding Stoks parameters $\xi^f_i$ read
\begin{eqnarray}
\xi^f_i={\rm Sp}(\rho^f\sigma_i)~,
\end{eqnarray}
where $\sigma_i$ is the Pauli matrix.
Thus, the parameter $\xi^f_3$ is equal to the asymmetry
\begin{eqnarray}
  \label{06}
\xi^f_3={\cal A}\equiv\frac{|M(1)|^2 - |M(2)|^2}{|M(1)|^2 + |M(2)|^2}~.
\end{eqnarray}
The sign of ${\cal A}$, plus or minus,
indicates the direction of $\gamma'$-polarization
relative to the ${e}'_{1}$ or ${e'}_2$ axes, respectively.
The degree of polarization ${\cal{P}}_{12}$ relative to axes 1 and 2
is related to the asymmetry as
\begin{eqnarray}
  \label{066}
 {\cal P}_{12}=\frac{1\pm{\cal A}}{2}~.
\end{eqnarray}

Our study is close to previous work~\cite{Serbo2004}
and recent~\cite{Seipt:2020diz}. Compared to~\cite{Serbo2004},
we now pay special attention to the kinematics of the LUXE
and other future experiments and concentrate on the photon polarization
in a wide region of $\xi$.
We have analyzed the impact of a finite number of
oscillations in the pulse describing the background field.
The difference from~\cite{Seipt:2020diz}
lies in some technical details: such as the choice of polarization axes,
the use of the asymmetry ${\cal A}$ as an important observable, and
the difference in calculation of transition amplitudes for the finite
e.m. pulse discussed in text.
In addition, we pay special attention to the azimuthal angle
distributions of the cross sections and asymmetries, which is
important for determination of $\gamma'$ polarization.
Our study seems to be useful as an additional and
independent analysis of alignment and degrees of $\gamma'$
polarization, which seems to be in demand in connection with
planned  experiments on electron-positron pair
production in electron-laser interactions.

Methodologically, this is close to~\cite{TK2020},
where the crossed nlBW process with linearly polarized photons
in the initial state in a wide range of e.m. strength is considered.

Our work is organized as follows.
Section II provides the basic formulas for cross
sections and asymmetries.
The cases $\xi\le 1$ and $\xi\gg1$ are discussed in
Sections III and IV, respectively.
Our summary is given in Section V.

\section{Cross sections and asymmetries}
As mentioned above, we consider nlCo as a process
involving a large number of laser photons simultaneously.
Our consideration is based on the Furry picture.
Here we present the basic formulas for cross sections and asymmetry,
which are used below.
We analyze them as the functions of the frequency
$\omega'$ and azimuthal angle $\varphi$ over a wide range of $\xi$.
The energy of the incoming electron $E_e=16.5$~GeV
and frequency of the optical laser pulse $\omega=1.55$~eV with
$\chi=0.196\,\xi$
are chosen according to the assumed values
in the LUXE experiment~\cite{LUXE_exp}.

For simplicity and without loss of generality,
in this section, we use the plane wave approximation
where the background field has a form of an monochromatic
plane wave (an infinite plane wave)
$A_\mu=a_\mu\cos\phi$ with $\phi=k\cdot x$ and
$a_\mu=(0,\xi m/|e|,0,0)$. The effect of a finite pulse with
a limited number of oscillations in the pulse
is discussed in Sect.~III~C.

The transition matrix element reads
\begin{eqnarray}
  S=
  \frac{i}{\sqrt{2\omega' 2q_0 2q'_0}}\sum\limits_{\ell=1}^{\infty}
    M_{\ell}(2\pi)^4\delta^{4} (q+\ell k-q'-k')~,
    \label{II001}
\end{eqnarray}
where $\ell$ is the number of {laser} beam
photons involved in the process,
and
\begin{eqnarray}
  &&M_{\ell}=\bar u_{p'}[{\fs{e}'}^*A_0(\ell)
  +(\frac{e}{2k\cdot p}{\fs{e}'}^*\fsk\fs{a}
  +\frac{e}{2k\cdot p'}\fs{a}\fsk {\fs{e}'}^*)A_1(\ell)
  \nonumber\\
  &&\qquad\qquad
  -\frac{e^2a^2(k\cdot {e'}^*)}{2k\cdot p\,k\cdot p'}\fsk A_2(\ell)]u_p,
 \label{II002} 
\end{eqnarray}
where $u_p$ is the Dirac spinor, normalized as $(\bar uu)=2m$. The
basic functions $A_m(\ell)\equiv A_m(\ell,\alpha',\beta)$
are determined as~\cite{RitusGroup} 
 \begin{eqnarray}
 A_m(\ell\alpha'\beta)=\frac{1}{2\pi}\int\limits_{-\pi}^{\pi} d\phi
 \cos^m(\phi)\,\,{\rm e}^{i\ell\phi - i\alpha'\sin\phi + i\beta\sin2\phi}
\label{II5555}
 \end{eqnarray} 
 with
\begin{eqnarray}
  \alpha'&=&z\cos\varphi,\quad\beta=\xi^3u/8\chi,\quad u=kk'/kp'~,
\label{ab6}\\
&&\nonumber\\
  z&=&\frac{2\ell\xi}{\sqrt{1+\xi^2/2}}
  \sqrt{\frac{u}{u_l}(1-\frac{u}{u_l})},\,\,
  u_l=\frac{2l\chi}{\xi(1+\xi^2/2)}.\nonumber\\
 \label{II6} 
\end{eqnarray}
The four vectors
$e'_i$ are chosen as~\cite{LL4}
\begin{eqnarray}
  e'_i=(0,\mathbf{e}'_i)~,
  \label{077}
\end{eqnarray}  
where $\mathbf{e}'_{1,2}$ are mutually orthogonal
and orthogonal to $\mathbf{k}'$: $\mathbf{e}'_{1,2},\perp\mathbf{k}'$.
The axes $\mathbf{e}'_{i}$ are selected  according
to~\cite{Akhiezer_Berestetsky}
\begin{eqnarray}
  {\mathbf e}'_1=\frac{[\mathbf k, \mathbf k']}{|[\mathbf k, \mathbf k']|},
  \qquad
  {\mathbf e}'_2=\frac{[\mathbf k', {\mathbf e'}_1]}{|\mathbf k'|}~,
 \label{07} 
\end{eqnarray}
leading to
\begin{eqnarray}
{\mathbf e}'_1&=&-{\mathbf x}\sin\varphi + {\mathbf y}\cos\varphi\nonumber\\
{\mathbf e}'_2&=&-{\mathbf x}\cos\theta\cos\varphi
      -{\mathbf y}\cos\theta\sin\varphi +{\mathbf z}\sin\theta~.
 \label{08} 
\end{eqnarray}

The physical meaning of ${\mathbf e}'_i$
becomes clear at backward scattering with $\cos\theta=-1$.
In a coplanar reference frame with $\varphi=0$, ${\mathbf e}'_{1}$
is parallel to the $\mathbf y$ axis, i.e. perpendicular
to the {laser} beam polarization, and
${\mathbf e}'_{2}$ is parallel to the $\mathbf x$ axis
or parallel to the {laser} beam polarization.
This means that positive (negative)
sign of ${\cal A}$ corresponds to the alignment of $\gamma'$
polarization perpendicular (parallel) to the {laser} beam polarization.
In the reference frame with $\varphi=\pi/2$,
the positive (negative) sign of ${\cal A}$ corresponds
to alignment of $\gamma'$ polarization parallel (perpendicular)
to the {laser} beam polarization.

{The choice of ${\mathbf e}'_i$ in
the form of (\ref{07})~\cite{Akhiezer_Berestetsky}
 seems to be attractive for its clarity
and physical transparency.
However, Refs.~\cite{Seipt:2020diz} and ~\cite{Rumyantsev}
use expressions for  ${\mathbf e}'_i$  different from (\ref{07}).
On the other hand, in the exact QED evaluation of Compton scattering
in a strong magnetic field performed in~\cite{Mushtukov},
${\mathbf e}'_i$ is chosen in the form of~(\ref{08}).
In coplanar geometry with $\varphi=0$ (which gives the dominant contribution
to the total cross section) all variants are equivalent.
Some differences between different variants may appear on
finite $\varphi$, which can be the subject of experimental study of this issue.}\\

The yield of $\gamma'$ production for the fixed polarization is equal to
the square of the sum of the matrix elements $\sum |M_\ell|^2/2VT$ multiplied
by the phase space factor of outgoing particles.
The yield expressed in $\varphi$ and $\omega'$ is calculated
by integrating the phase space factor over $d^3q'$
\begin{eqnarray}
\delta^4(\ell k + q-q'-k')\frac{d^3q'd^3k'}{q'_0k'_0}  
\to \frac{d\varphi d\omega'}{|\vec{q} - \ell k_0|}~.
\nonumber
\end{eqnarray}
{The corresponding cross sections are determined
by multiplying the yield by
the flux factor
${q_0}\omega/{p\cdot k}$
$\times$
$\rho^{-1}_\gamma$, where $\rho_\gamma={m^2\xi^2\omega}N_0/{4\pi\alpha}$
is the electromagnetic laser field energy density
divided by $\omega$:
 $\langle{\it E_{e.m.}}\rangle/\omega$.
For monochromatic background the flux parameter $N_0=1/2$.
For a pulse with a “finite” number of oscillations,
$N_0$ depends on the wave envelope (cf. Sect.~III~C).}

In case of monochromatic background the matrix element squared 
for the fixed  ${\mathbf e'}_1$ and  ${\mathbf e'}_2$ 
is calculated in a standard way, using the gauge
transformation~\cite{Ritus-79,Greiner}
$e'\to \bar e'_i=e'_i- k'(e'_i\cdot k)/(k'\cdot k)$
and relation~\cite{RitusGroup}
\begin{eqnarray}
  \alpha' A_1=(\ell -\beta)A_0 + 4\beta A_2.
  \label{relations_for_A}
\end{eqnarray}  
As a result, the "partial'' cross sections $d^2\sigma_{1,2}$, averaged and summed
over the initial and final electron spin projections, respectively,  read
\begin{eqnarray}
{d^2 \sigma_{1}}
&=&\frac{2\alpha^2d\varphi d\omega'}{\xi\chi m^2~N_0}
\sum\limits_{\ell=\ell_{\rm min}}^{\infty}
\,\frac{1}{|\vec{q}-l\omega|}
\, [\xi^2~A_1^2\sin^2\varphi\nonumber\\
  &+&\xi^2\frac{u^2}{4(1+u)}\left(A_1^2-A_0A_2\right)]~,
\label{II6_}\\
{d^2 \sigma_{2}}
&=&\frac{2\alpha^2 d\varphi d\omega'}{\xi \chi m^2~N_0}
\sum\limits_{\ell=\ell_{\rm min}}^{\infty}\,\frac{1}{|\vec{q}-l\omega|}
\, [-A_0^2 -\xi^2\,A_1^2\sin^2\varphi\nonumber\\
 &+&\xi^2 \left(1+\frac{u^2}{4(1+u)}\left(A_1^2-A_0A_2\right)\right)]~,
\label{II7}
\end{eqnarray}

The sum $d^2\sigma\equiv d^2\sigma_1+d^2\sigma_2$ (for monochromatic pulse)
is the known expression for the unpolarized cross section
\begin{eqnarray}
{d^2 \sigma}
&=&\frac{4\alpha^2 d\varphi d\omega'}{\xi m^2\chi}
\sum\limits_{\ell=\ell_{\rm min}}^{\infty}\,\frac{1}{|\vec{q}-l\omega|}
\, [-A_0^2 \nonumber\\
&+&\xi^2 \left(1+\frac{u^2}{2(1+u)}\left(A_1^2-A_0A_2\right)\right)]~,
\label{IItotal}
\end{eqnarray}
where $\ell_{\rm min}=u\xi(1+\xi^2/2)/2\chi$.

The difference in $d^2\sigma_{1,2}$ leads to 
asymmetry (cf.~Eq.~(\ref{06}))
\begin{eqnarray}
  {\cal A}(\varphi,\omega')=\frac{d^2\sigma_1 - d^2\sigma_2}
  {d^2\sigma}~.
\label{II9}
\end{eqnarray}
The sign of ${\cal A}$, plus or minus, is related to the direction
of $\gamma'$-polarization relative
to the axes $\mathbf{e}'_{1}$ or $\mathbf{e}'_{2}$, respectively.
The degree of polarization is determined according to Eq.~(\ref{066}).

Since the asymmetry is defined for any pair
of $\varphi$ and $\omega'$, 
the average asymmetry $\langle{\cal A}\rangle_\varphi$ can be
introduced
\begin{eqnarray}
\langle{\cal A}(\omega')\rangle_{\varphi}=\frac{1}{2\pi}\int\limits_0^{2\pi}
  d\varphi\,{\cal A}(\varphi,\omega')~.
\label{II11}  
\end{eqnarray}
It has a physical meaning and can be studied experimentally.
We emphasize that asymmetry $\cal A$ is equal to the $\gamma'$
spin variable $\xi_3^f$.

Finally note, that $\cos\varphi$  changes sign upon transformation
$\varphi\to\varphi + \pi$. Using the symmetry property of
the basis functions~(\ref{II5555})
$A_m(l,\alpha',\beta)$ with $\alpha'=z\cos\varphi$,
$A_m(l,-\alpha',\beta)=(-1)^{l+m}A_m(l,\alpha',\beta)$)~\cite{Ritus-79},
one can find that the bi-linear combinations
$A_0^2$, $A_1^2$, $A_0A_2$, and $A_1^2\sin^2\varphi$ 
are even functions
of $\cos\varphi$, which leads to the symmetry relation
\begin{eqnarray}
  d^2\sigma_i(\varphi)=d^2\sigma_i(\varphi+\pi)~.
\label{symmetry}  
\end{eqnarray}  
\section{Finite field strength (${\mathbf{\xi}}\le 1$)}
\subsection{Low field strength with ${\mathbf{\xi}}\ll 1$}

For methodological purposes, we consider
here a monochromatic pulse with a weak field
strength with $\xi\ll 1$, where
the lowest harmonic with $l=1$ dominates.
The functions $A_m(l=1,\alpha',\beta)$ for small values of
the parameters $\alpha'\sim\xi$, $\beta\sim\xi^2$ read
\begin{eqnarray}
  A_0\simeq \frac{\alpha'}{2}(1+\frac{\beta}{2}),\,
  A_1\simeq\frac12 -\frac{{\alpha'}^2}{16}-\frac{\beta}{4},\,
  A_2\simeq\frac{\alpha'}{8}(1+\beta).\nonumber
\end{eqnarray}
In the frame  with $\vec p=0$ the total cross section is
\begin{eqnarray}
  \frac{d\sigma}{d\Omega}
  =\sigma_0
  \left(\frac{\omega'}{\omega}\right)^2
  \left(\frac{\omega'}{\omega}+\frac{\omega}{\omega'}
  -2\sin^2\theta\cos^2\varphi\right)~, 
  \label{II1111}  
\end{eqnarray}
where $\sigma_0=\alpha^2/2m^2$ and the identity
$d\omega'/|\vec{q}-\ell\omega|={\omega'}^2d\cos\theta/ \ell k\cdot p$
is used. Averaging over $\varphi$ leads to the well-known Klein-Nishina
cross section.

The partial cross sections read
\begin{eqnarray}
  \frac{d\sigma_1}{d\Omega}=\sigma_0\left(\frac{\omega'}{\omega}\right)^2
\left(\frac12\left(\frac{\omega'}{\omega} + \frac{\omega}{\omega'}\right)
+ 2\sin^2\varphi -1\right)~,
\label{II122}
\end{eqnarray}
\begin{eqnarray}
\frac{d\sigma_2}{d\Omega}&=&\sigma_0\left(\frac{\omega'}{\omega}\right)^2
  \left(\frac12\left(\frac{\omega'}{\omega} + \frac{\omega}{\omega'}\right)
  - 2\sin^2\varphi+1\right.\nonumber\\
  &-& \left. 2\sin^2\theta\cos^2\varphi\right)~.
  \label{II133}
\end{eqnarray}
Equations~(\ref{II122}), (\ref{II133}) may be written in a compact form
known from textbooks~\cite{Akhiezer_Berestetsky,LL4,Greiner}.
\begin{eqnarray}
\frac{d\sigma_i}{d\Omega}=\frac12\sigma_0\left(\frac{\omega'}{\omega}\right)^2
  \left(\frac{\omega'}{\omega} + \frac{\omega}{\omega'}
  - 2 + 4(\mathbf{e}_0\mathbf{e}'_i)^2\right)~,
  \label{II15}
\end{eqnarray}
where $\mathbf{e}_0=\mathbf{x}$ and $\mathbf{e}'_i$ are the polarization vectors
of $\gamma$ and $\gamma'$, respectively.

Equations~(\ref{II122}), (\ref{II133}) lead to the asymmetry
\begin{eqnarray}
{\cal A}(\varphi,\omega')  =
  2\frac{2\sin^2\varphi +\sin^2\theta\cos^2\varphi -1}
  {\frac{\omega'}{\omega} + \frac{\omega}{\omega'}
 - 2\sin^2\theta\cos^2\varphi}.   
  \label{II14}  
\end{eqnarray}

In coplanar geometry with $\varphi=0$, the asymmetry is proportional
to $-\cos^2\theta$, i.e. negative.
This means that the polarization of the recoil photon is directed
along the $\mathbf{e}'_2$ axis, or along the {laser} beam polarization.
In the frame with $\varphi=\pi/2$, the asymmetry is positive,
it is directed along the $\mathbf{e}_1$ axis, and is again
parallel to the {laser} beam polarization.
In the general case, the asymmetry of ${\cal A}$ for separate
$\ell>1$ depends on the value of $\varphi$ and the parity of the harmonics.

\begin{figure}[h]
  \includegraphics[width=0.6\columnwidth]{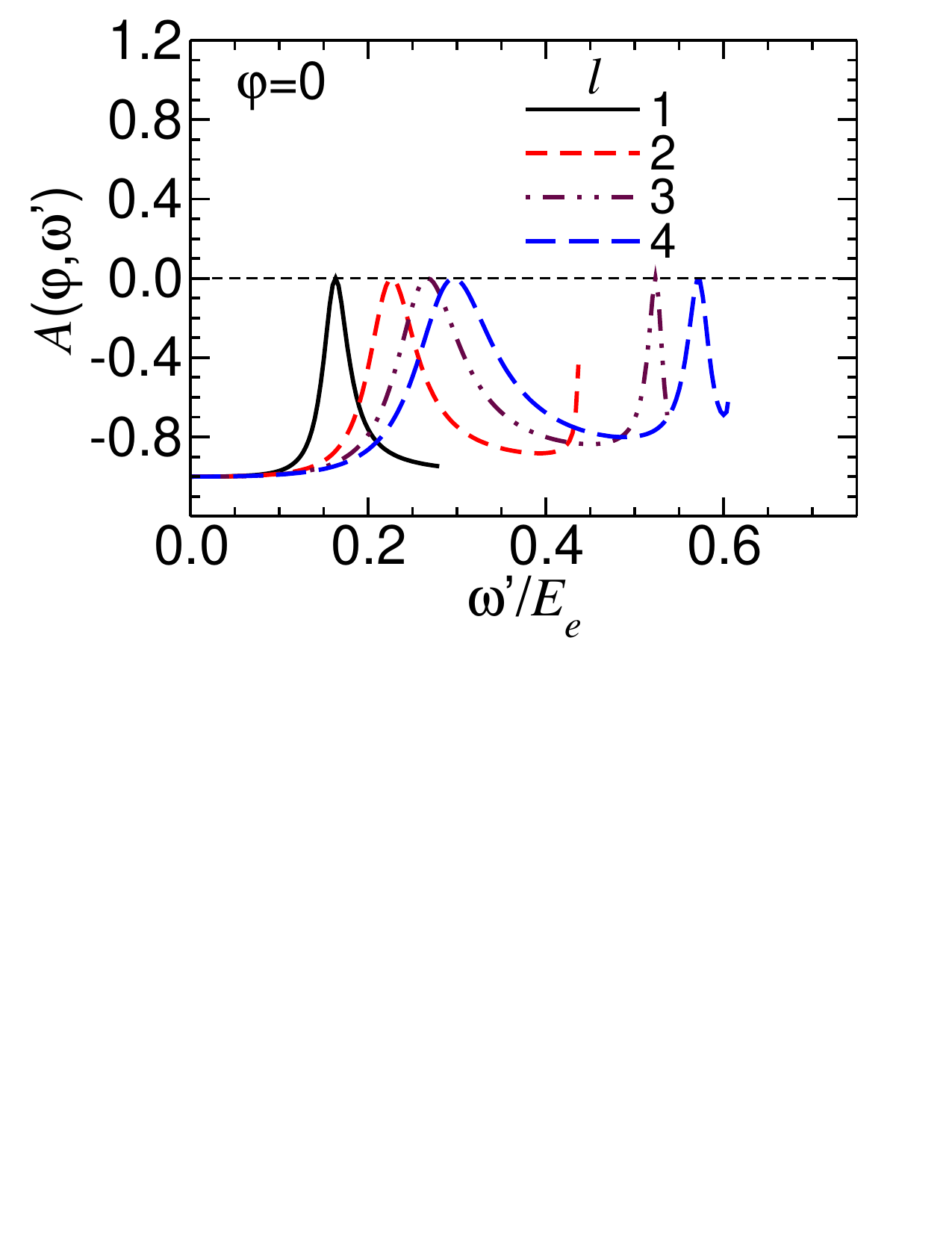}\\
  \vspace{-3.2cm}  
\includegraphics[width=0.6\columnwidth]{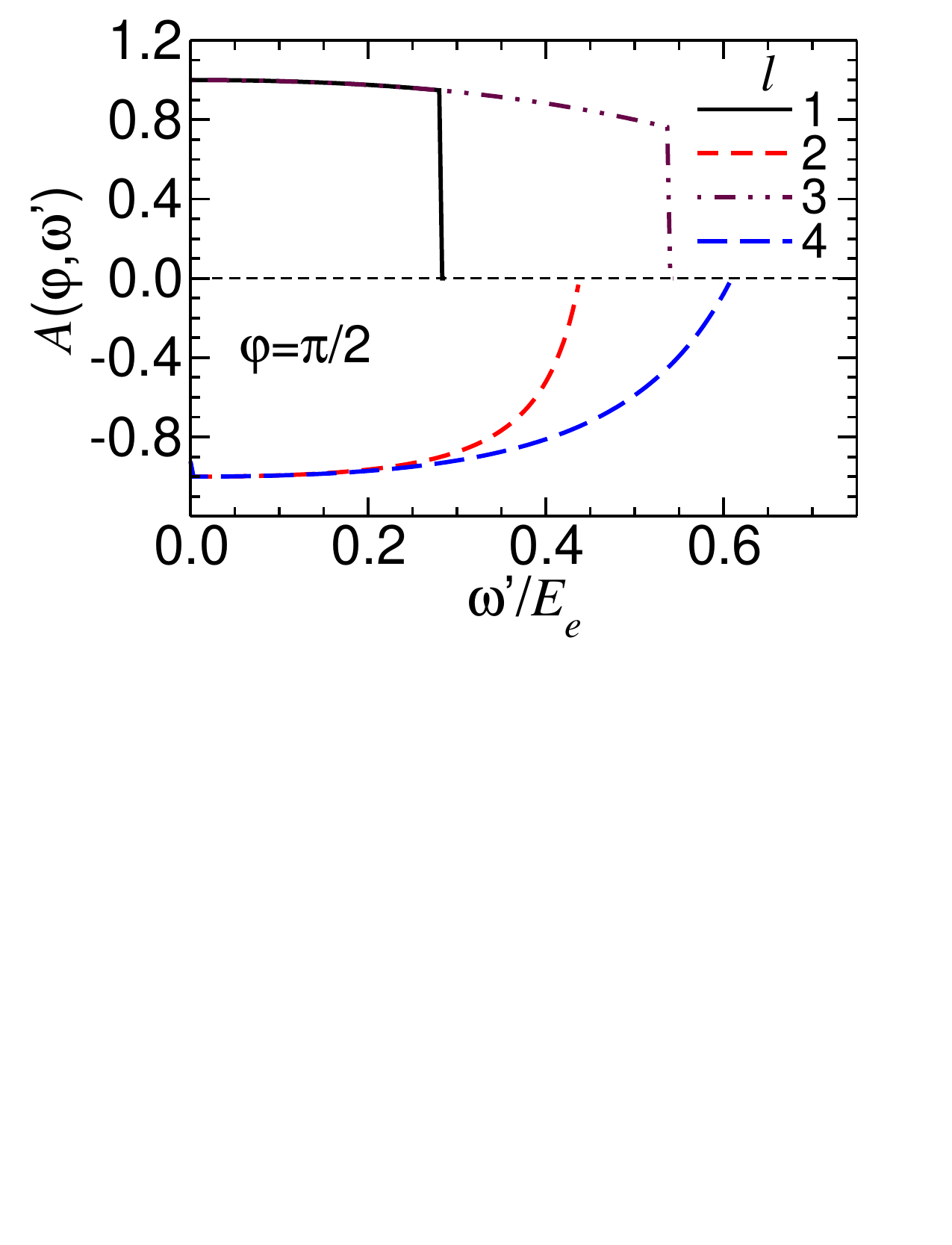}
\vspace{-3.6cm}
\caption{(Color online)
  The asymmetry ${\cal A}(\varphi,\omega')$ for monochromatic
  background as a function of $\omega'$
  for $\ell=1...4$ at $\varphi=0$
  and ${\pi}/{2}$ is shown in the top and bottom
  panels, respectively. The electron energy $E_e=16.5$~GeV,
  {laser} beam frequency $\omega=1.55$~eV, and $\xi=0.1$.
  \label{Fig:1}}
\end{figure}

For example, the asymmetry ${\cal A}(\varphi,\omega')$ for individual
harmonics $\ell=1...4$ at $\xi=0.1$ for $\varphi=0$ and $\pi /2$
as a function of $\omega'/E_e$
is shown in Fig.~\ref{Fig:1} in the top and bottom panels, respectively.
{This analysis (for separate harmonics)
  is made mainly for methodological purposes and comparison
with Ref.~\cite{Serbo2004}.}

For $\varphi=0$ (coplanar frame), the asymmetry is negative for
all harmonics. For $\ell=1$ it has a bump-like behavior
with a bump position $\omega'/E_e\approx 0.17$.
For $\ell>1$, the number of bumps increases and
the position of the first bump shifts towards larger
values of $\omega'$.

In the case of $\varphi={\pi}/{2}$, the asymmetry is positive
and large for odd $\ell$ and negative for even $\ell$ (for separate harmonics!).
This result is consistent with \cite{Serbo2004}
(cf. Fig.~8).

\subsection{Monochromatic background with $\xi\le1$}
\begin{figure}[h]
  \includegraphics[width=0.5\columnwidth]{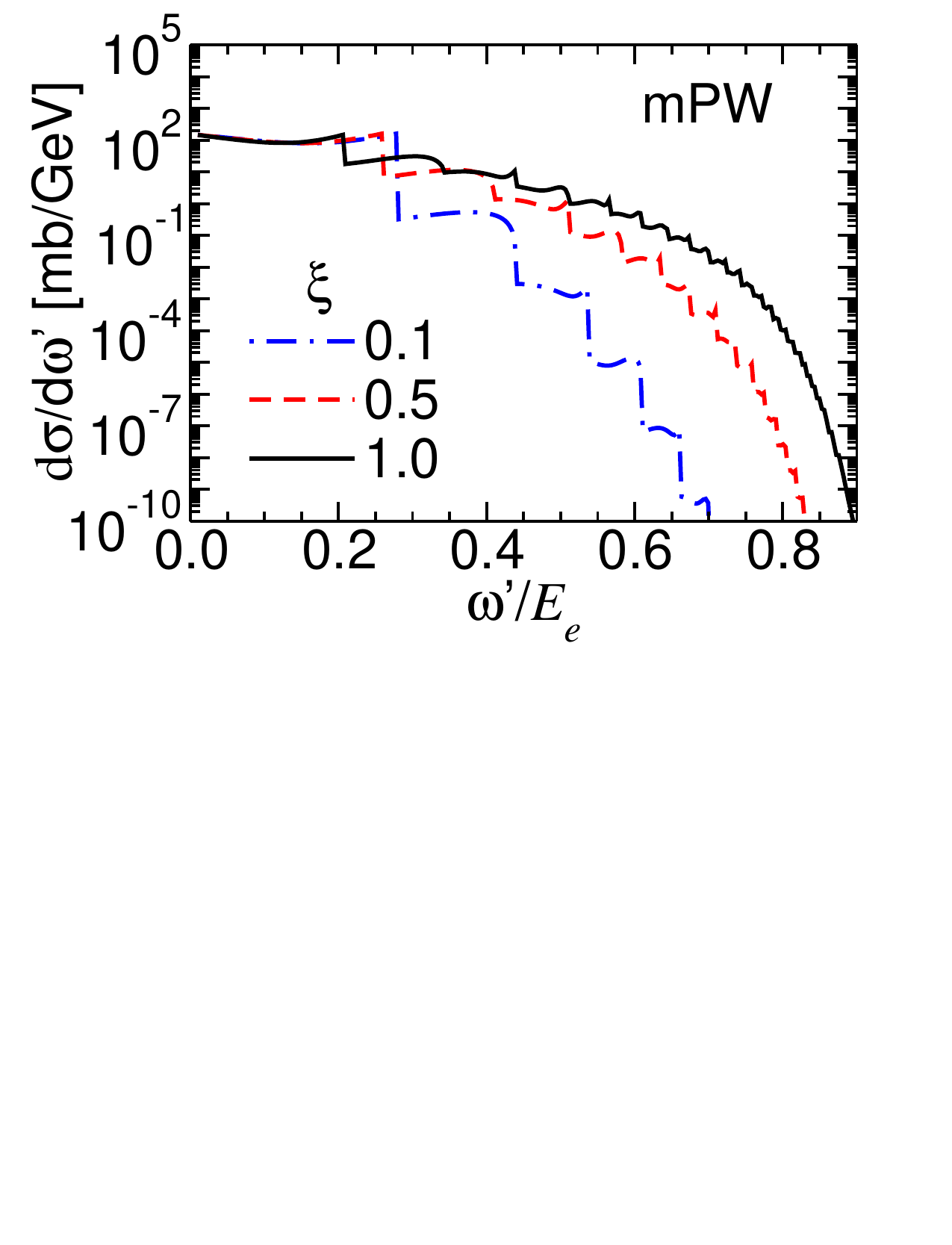}\\
\vspace*{-3.0cm}
\caption{\small{(Color online)
    The differential unpolarized cross sections
    $d\sigma/d\omega'$ (\ref{IItotal}) as functions
    of $\omega'$ for different  $\xi$ for monochromatic plane wave background.
\label{Fig:2}}}
\end{figure}

Consider now the result for the coherent sum of all harmonics
in (\ref{II6_}), (\ref{II7}) for the monochromatic plane wave background (mPW).
Here and below, the electron energy $E_e=16.5$ GeV
and the {laser} beam frequency $\omega=1.55$~eV are chosen
according to the LUXE kinematics~\cite{LUXE_exp}.
The unpolarized cross section $d\sigma/d\omega'$ (\ref{IItotal})
as a function of $\omega'$ is exhibited in Fig.~\ref{Fig:2}.
{The result is consistent with other similar
researches (see, for example, relevant references
in~\cite{TitovPEPAN}).
At small $\xi^2\ll1$, the cross sections have a step-like behavior
(cf. bottom curve in Fig.~\ref{Fig:2}),
where each new step corresponds to the inclusion of a new harmonic
as $\omega'\to \omega'_{\rm max}(\ell)$ with
\begin{eqnarray}
  \omega'_{\rm max}(\ell)=\frac{\ell k\cdot q}
         {q_0+|\vec q|\cos\theta +\ell\omega(1-\cos\theta)}~.
  \label{II17}  
\end{eqnarray}
The height of the steps is proportional to $\xi^2$.
It is explained by the properties of the basis functions $A_m$~[4],
and the presence of the factor $\xi^2$ in (15) and (16).
As $\xi$ increases, the relative height of the steps decreases and
at $\xi=1$ becomes insignificant (see the top curve in Fig.~2)
cf.~\cite{Titov2014,impact}.}

\begin{figure}[h]
 \includegraphics[width=0.49\columnwidth]{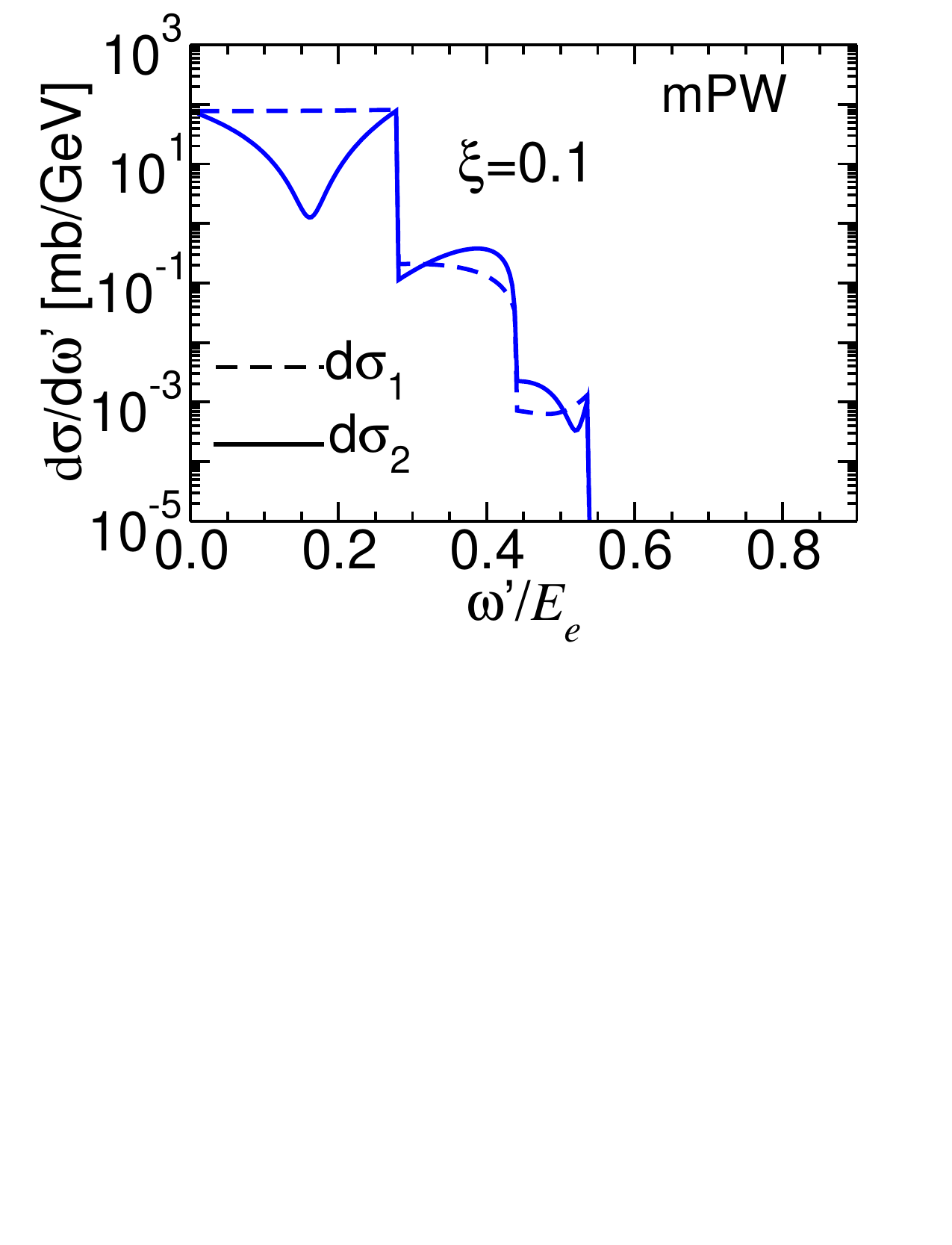}
  \includegraphics[width=0.49\columnwidth]{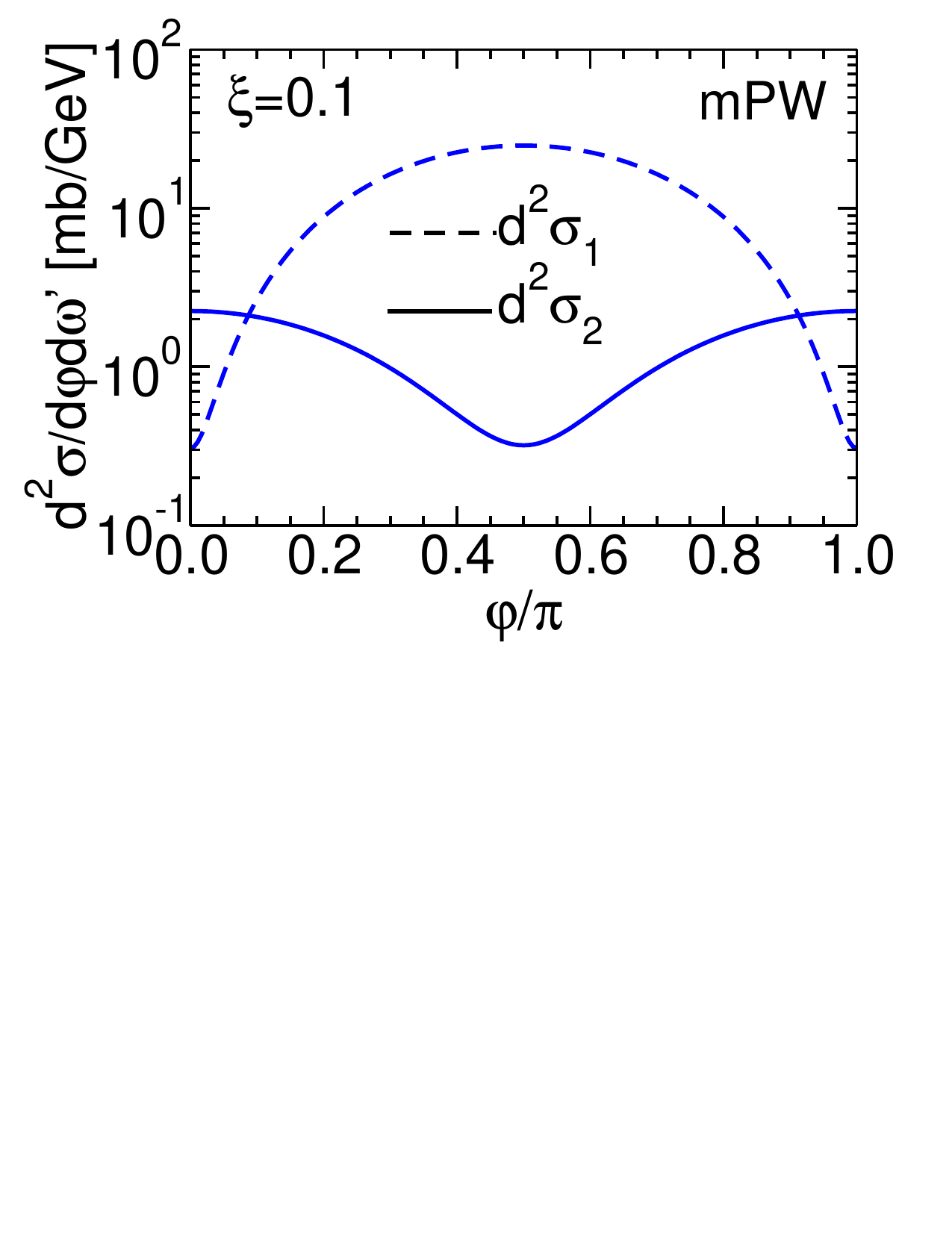}\\ 
  \vspace*{-2.6cm} 
  \includegraphics[width=0.49\columnwidth]{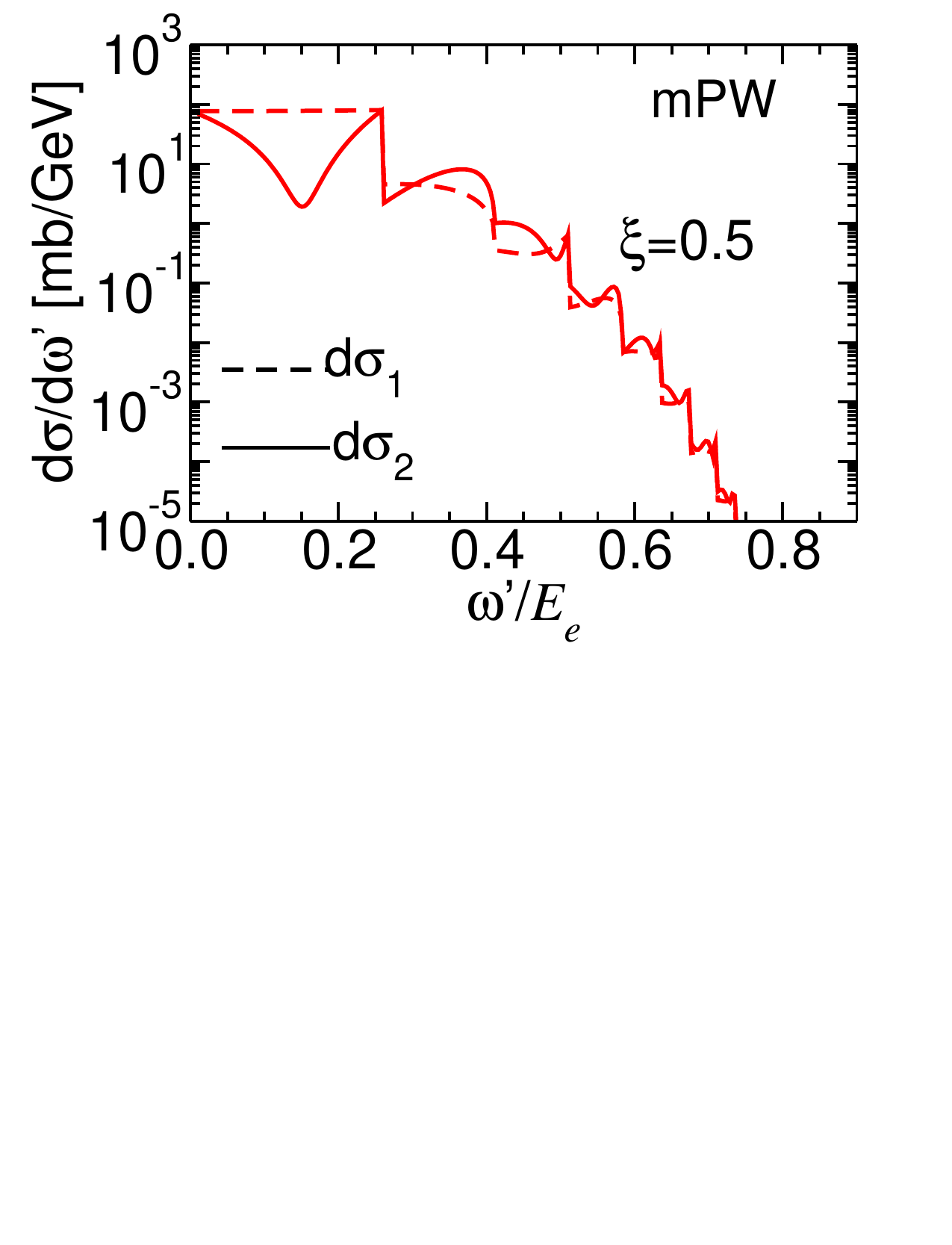}
  \includegraphics[width=0.49\columnwidth]{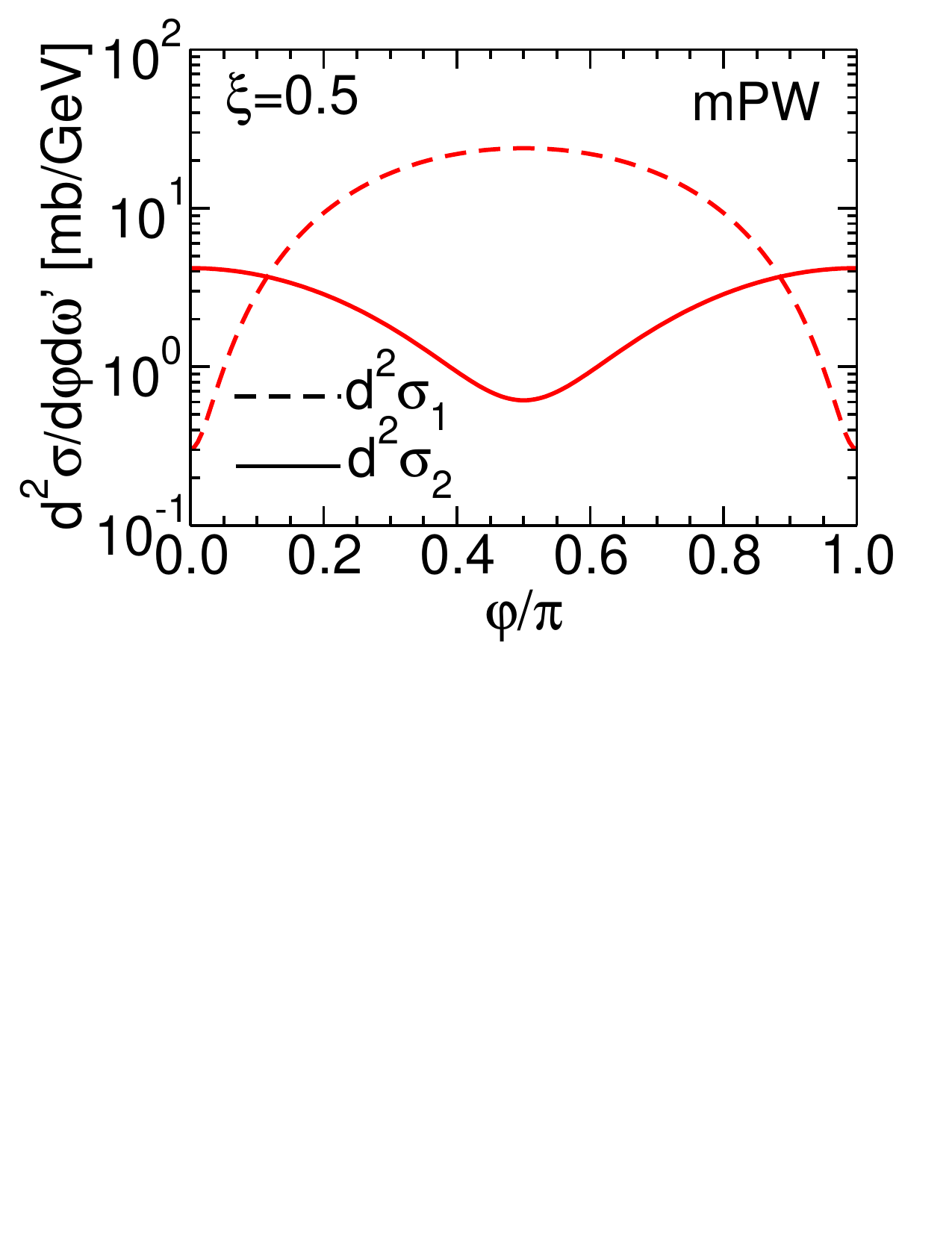}\\
  \vspace*{-2.6cm} 
  \includegraphics[width=0.49\columnwidth]{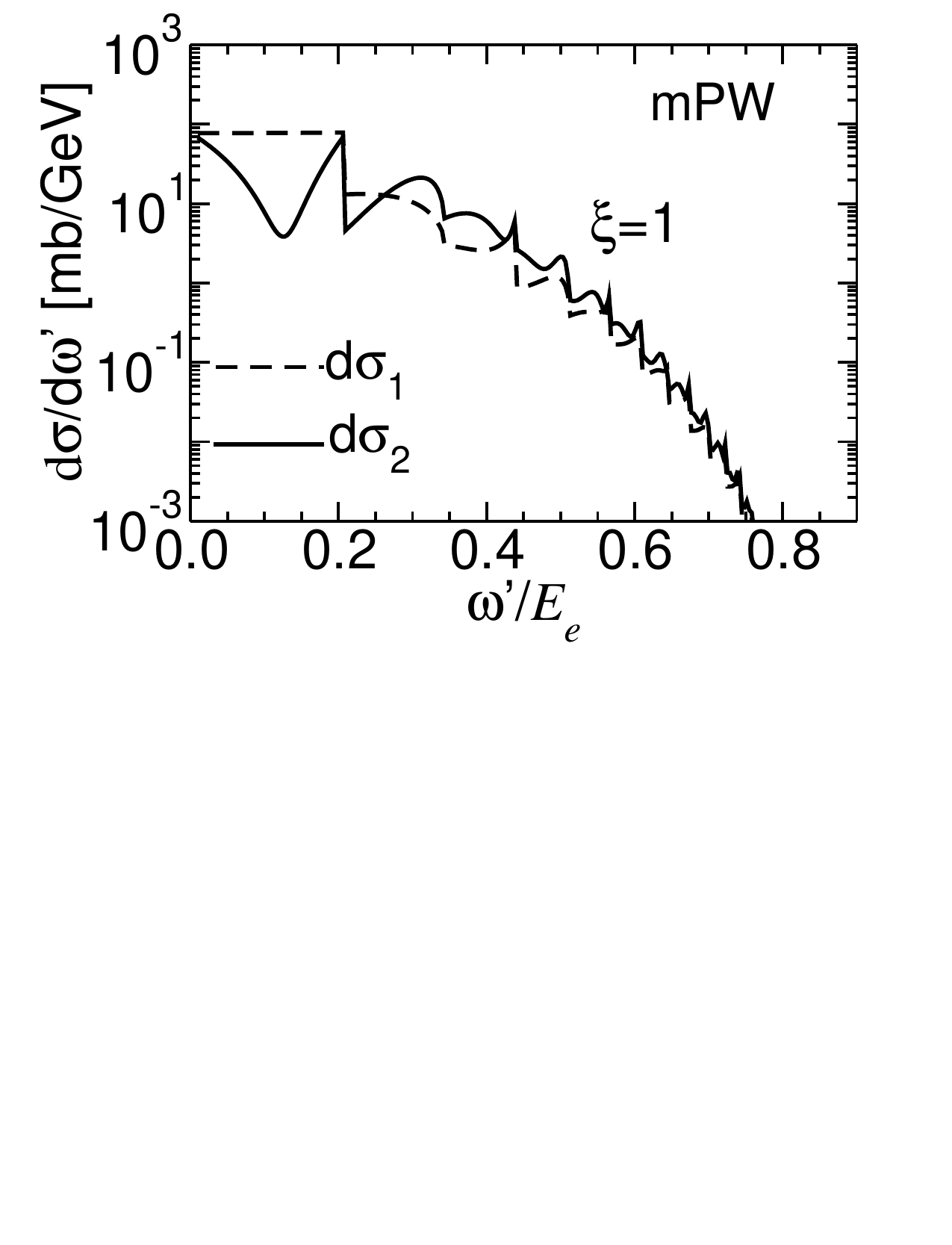}
  \includegraphics[width=0.49\columnwidth]{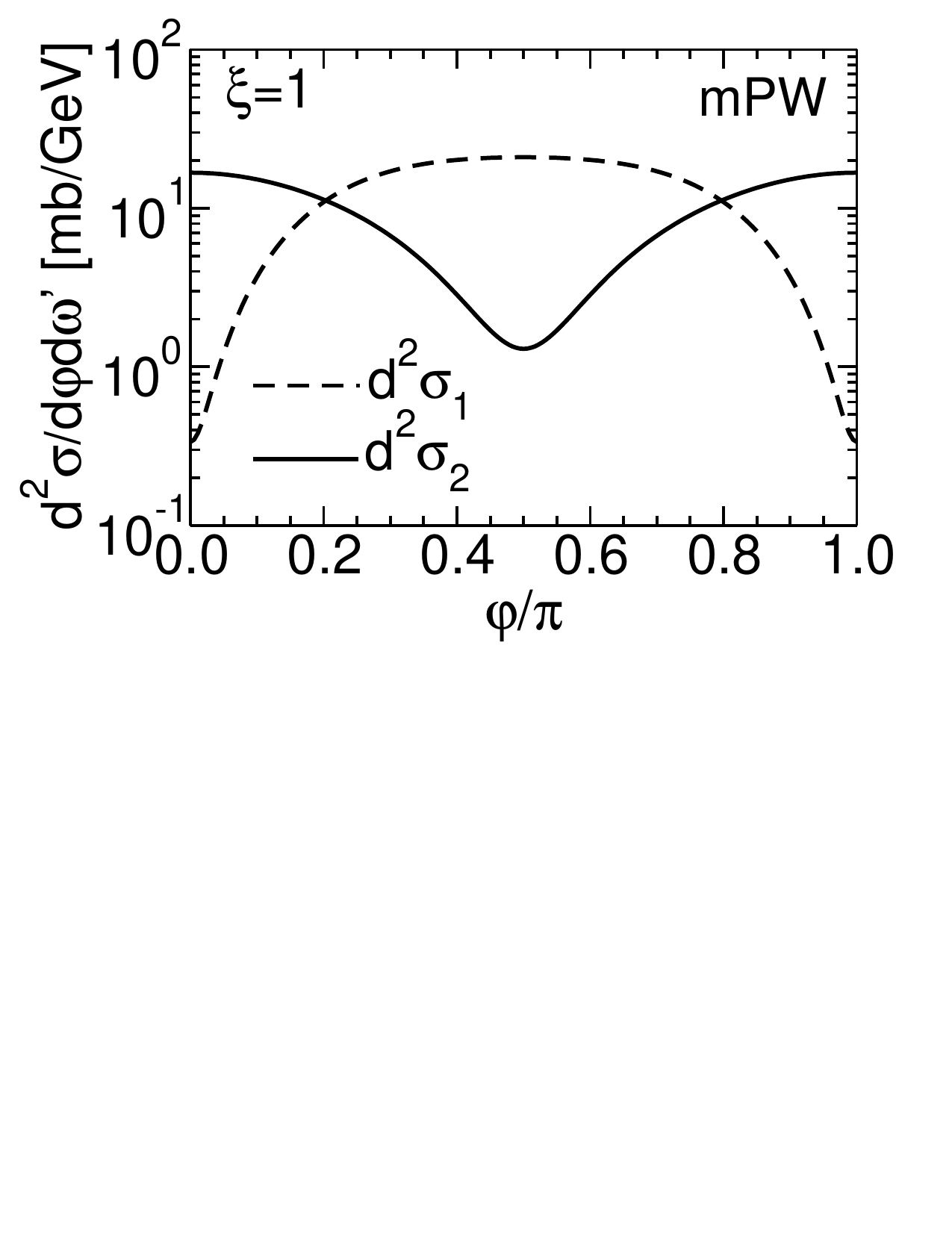}      
 \vspace*{-3.4cm}   
 \caption{\small{(Color online)
     The left panels show the partial cross sections $d\sigma_{1,2}/d\omega'$
     as functions of $\omega'$.     
     The right panels show the differential
     cross sections $d^2\sigma_{1,2}/d\varphi d\omega'$
     as functions of the azimuthal angle $\varphi$
     at  fixed $\omega'=0.2\,E_e$.
     Background is a  monochromatic plane wave.
     \label{Fig:3}}}
\end{figure}

The partial cross sections $d\sigma_{1,2}/d\omega'$ as functions of
$\omega'$ for different $\xi$ are shown
in the left panels of Fig.~\ref{Fig:3}.
In the region $\omega'<\omega'_1= \omega'_{\rm max}(\ell=1)$,
$d\sigma_2/d\omega'$ has a deep minimum.\\
The distributions of $d^2\sigma_i/d\varphi d\omega'$ over azimuthal angle
at fixed $\omega'=0.2\,E_e$, close to the position of the minimum
in $d\sigma_2/d\omega '$, are shown in the right panels of Fig.~\ref{Fig:3}.
Here we assume the symmetry property (\ref{symmetry}).
At $\varphi=0$ ($\pi$), $d^2\sigma_2\gg d^2\sigma_1$,
whereas at  $\varphi=\pi/2$, $d^2\sigma_1\gg d^2\sigma_2$, which results
in negative and positive asymmetries ${\cal A}(\varphi)$, for these two cases,
respectively.

 \begin{figure}[h]
 \includegraphics[width=0.49\columnwidth]{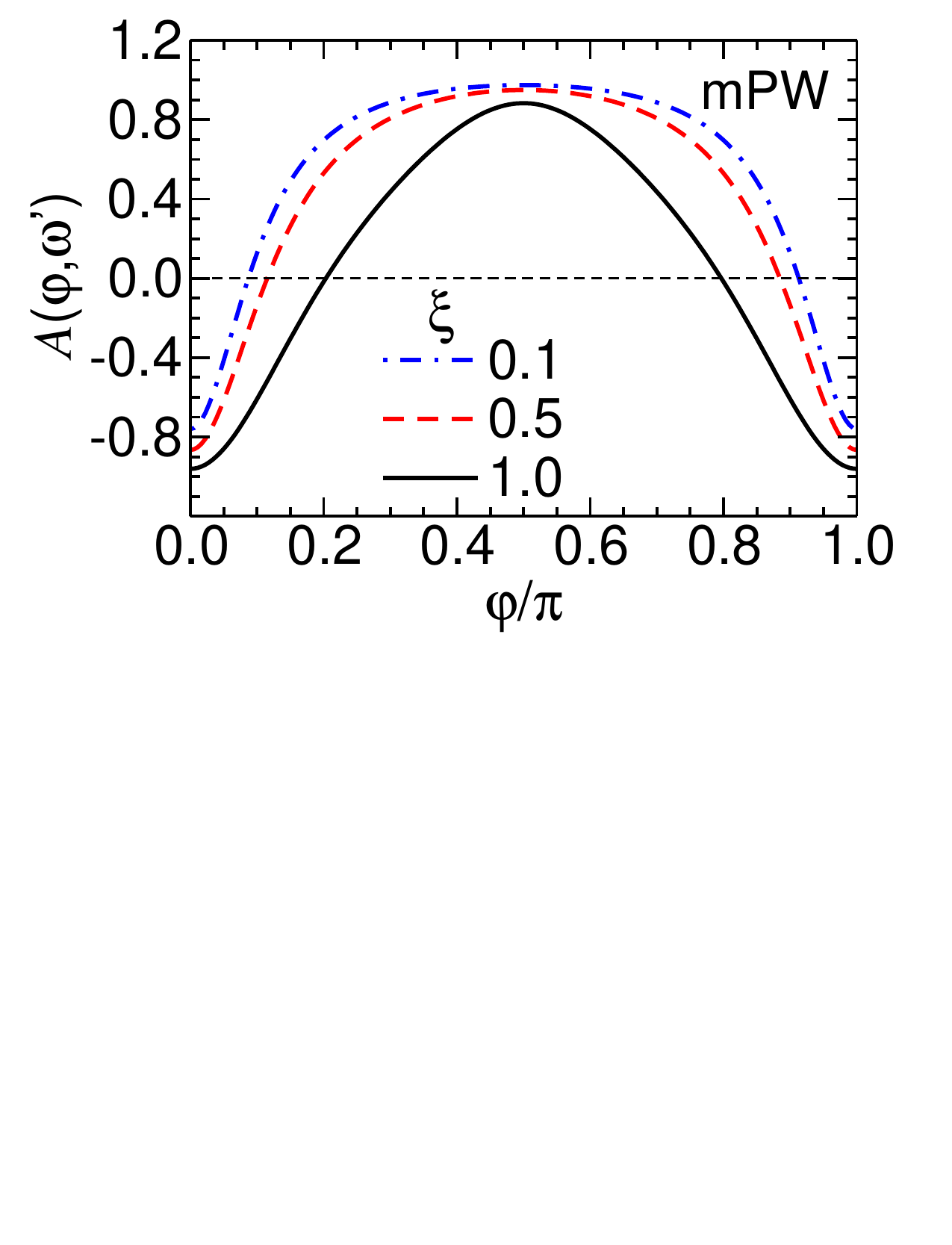}
 \includegraphics[width=0.49\columnwidth]{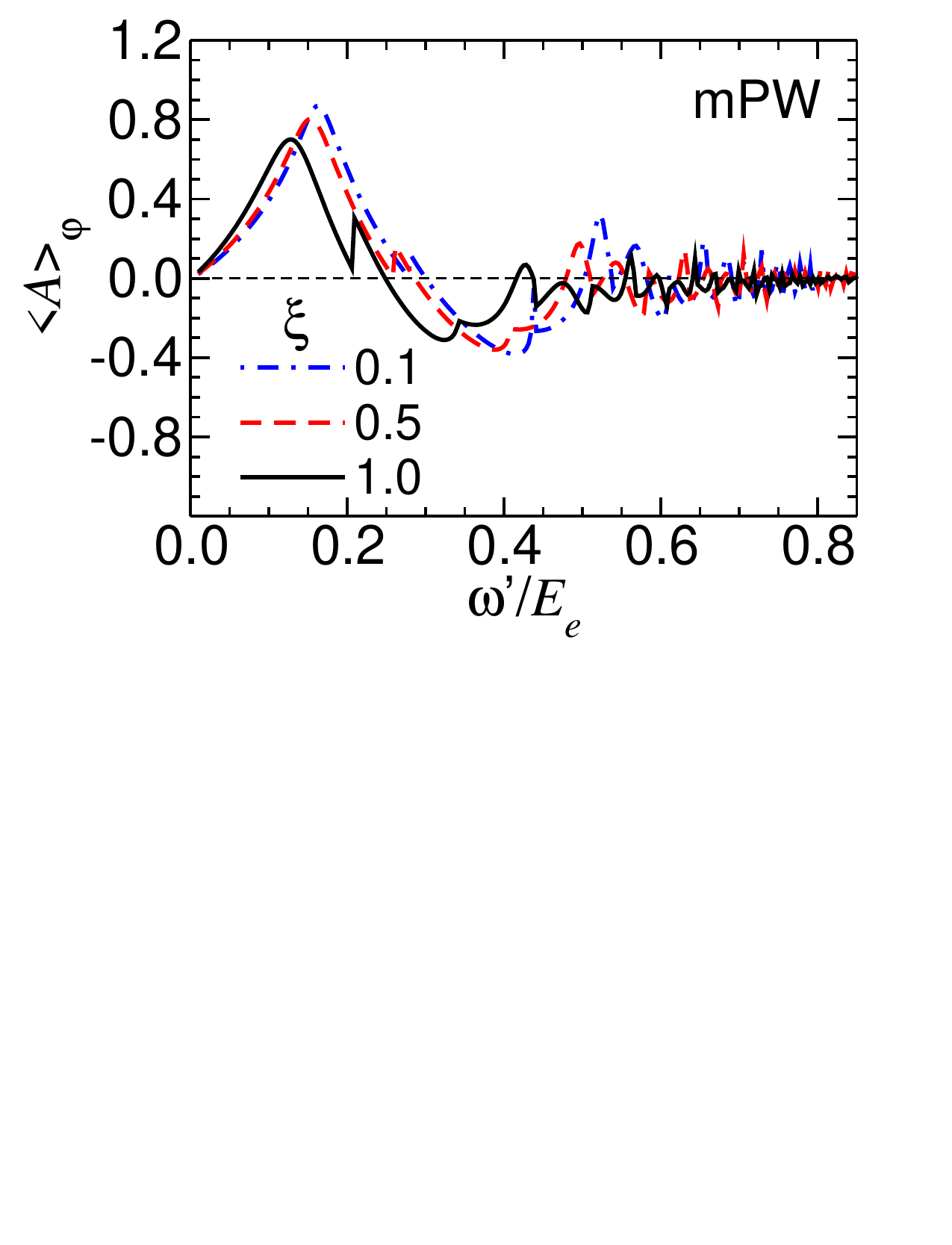} 
\vspace*{-3.4cm} 
 \caption{\small{(Color online)
     The left panel: The asymmetries as functions of $\varphi$
     at $\omega'=0.2\,E_e$ for different $\xi$.
     The right panel: The average asymmetries
     $\langle{\cal A}\rangle_\varphi$ as functions of $\omega'$.
      Background is a  monochromatic plane wave.
     \label{Fig:4}}}   
\end{figure}
The asimuthal angle dependence of ${\cal A}(\varphi,\omega')$
at fixed $\omega'=0.2E_e$ for different $\xi$ is exhibited
in the left panel of Fig.~\ref{Fig:4}. It is negative with a large absolute
value at $\varphi=0$ ($\pi$) and large and positive at $\varphi\simeq\pi/2$.
In both cases, polarization of $\gamma'$ is {\it along the  laser beam polarization}
with a degree of polarization ${\cal P}_{12}\simeq 0.9$.

The averaged asymmetry as a function of $\omega'$ is exhibited in the right panel of
Fig.~\ref{Fig:4}.
At small $\omega'$ it is positive with a bump around $\omega'/E_e\sim 0.2$,
then change a sign  to negative and vanishes at $\omega'\to E_e$.


\subsection{The finite pulse}
In this subsection, we study the effect of a
finite pulse, where the linearly polarized background field is
determined by the e.m. four-potential $A(\phi)=(0,{\mathbf A}(\phi))$ with
 \begin{eqnarray}
  \mathbf{A}(\phi) = f(\phi) \left[ \mathbf{a}\cos(\phi)\right]~,
  \label{Four_A}
 \end{eqnarray}
 where $\phi=k\cdot x$ is the invariant phase, and
 ${\mathbf a}={\mathbf x} m\xi/e$.
 For the sake of definiteness,
 the envelope function $f(\phi)$ is chosen as a hyperbolic secant:
$f(\phi)=1/[\cosh\phi/\Delta]$.
 The dimensionless quantity $\Delta$ is related to the pulse duration
 $2\Delta=2\pi N$, where $N$ has the meaning of the number of cycles in
 the laser pulse.
 It is related to the time duration of the pulse $\tau_N=2N/\omega$
(for the dependence of some observables on the envelope shape, see,
 for example \cite{TitovPEPAN,Titov2014}).

 \begin{figure}[ht]
\includegraphics[width=0.49\columnwidth]{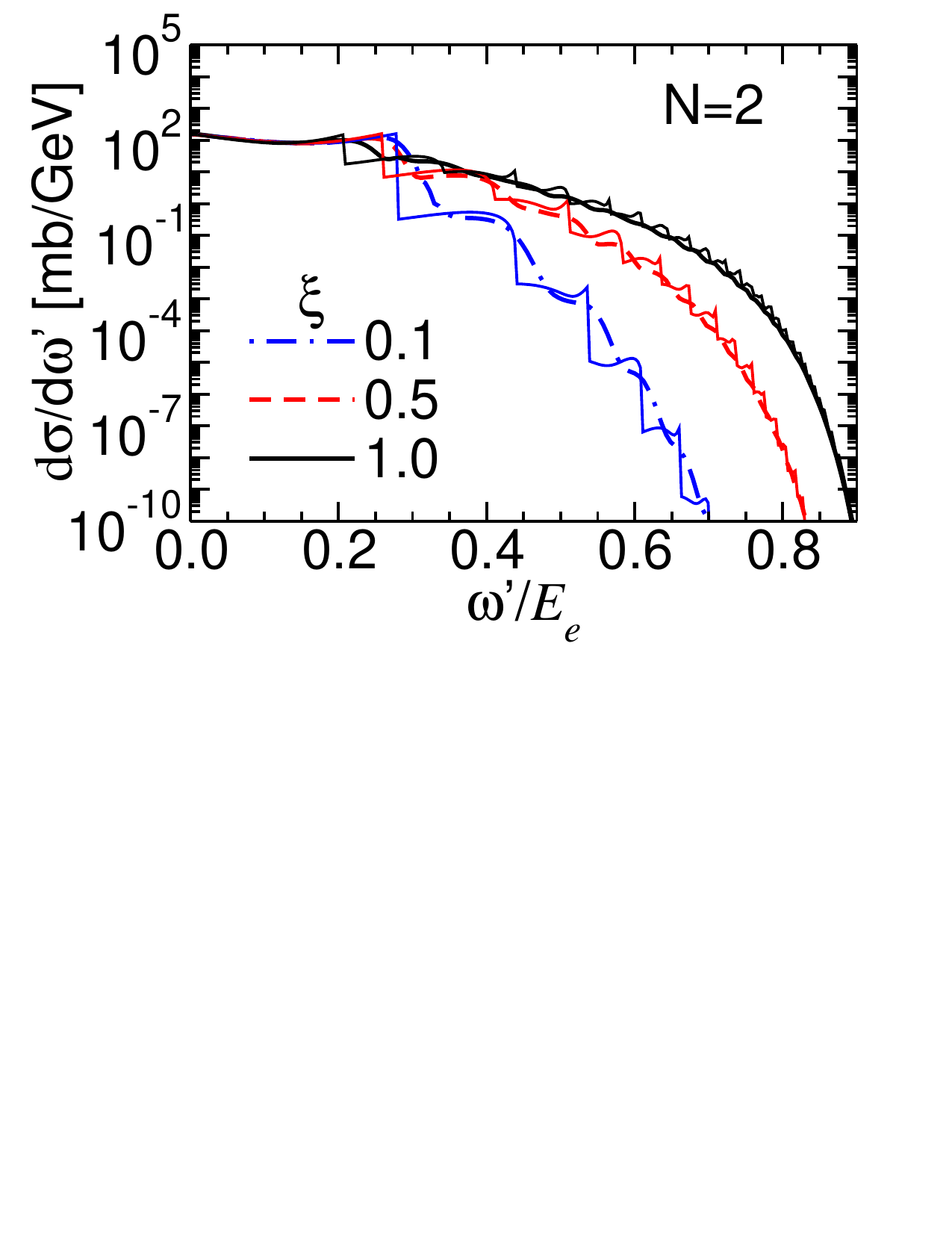}
\includegraphics[width=0.49\columnwidth]{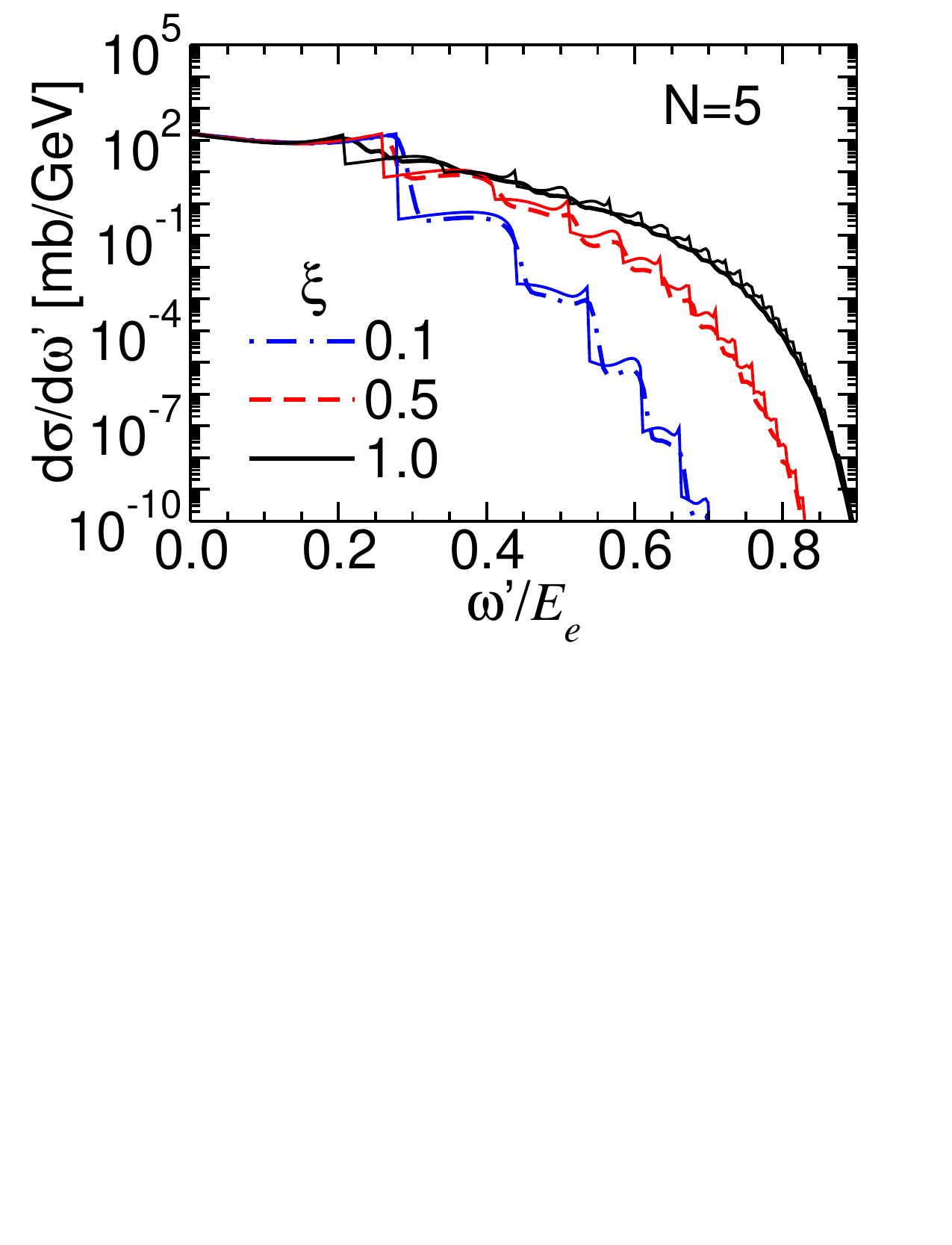}  
 \vspace*{-3.4cm}   
 \caption{\small{(Color online) 
     The differential unpolarized cross sections (\ref{III22})
     for different $\xi$. The left
     and right panels correspond to the number of oscillations
     in the pulse $N=2$ and 5, respectively.
     The corresponding result for monochromatic background is exhibited by the
     thin solid curves.
   \label{Fig:5}}}
 \end{figure}
 
  \begin{figure}[h]
  \includegraphics[width=0.49\columnwidth]{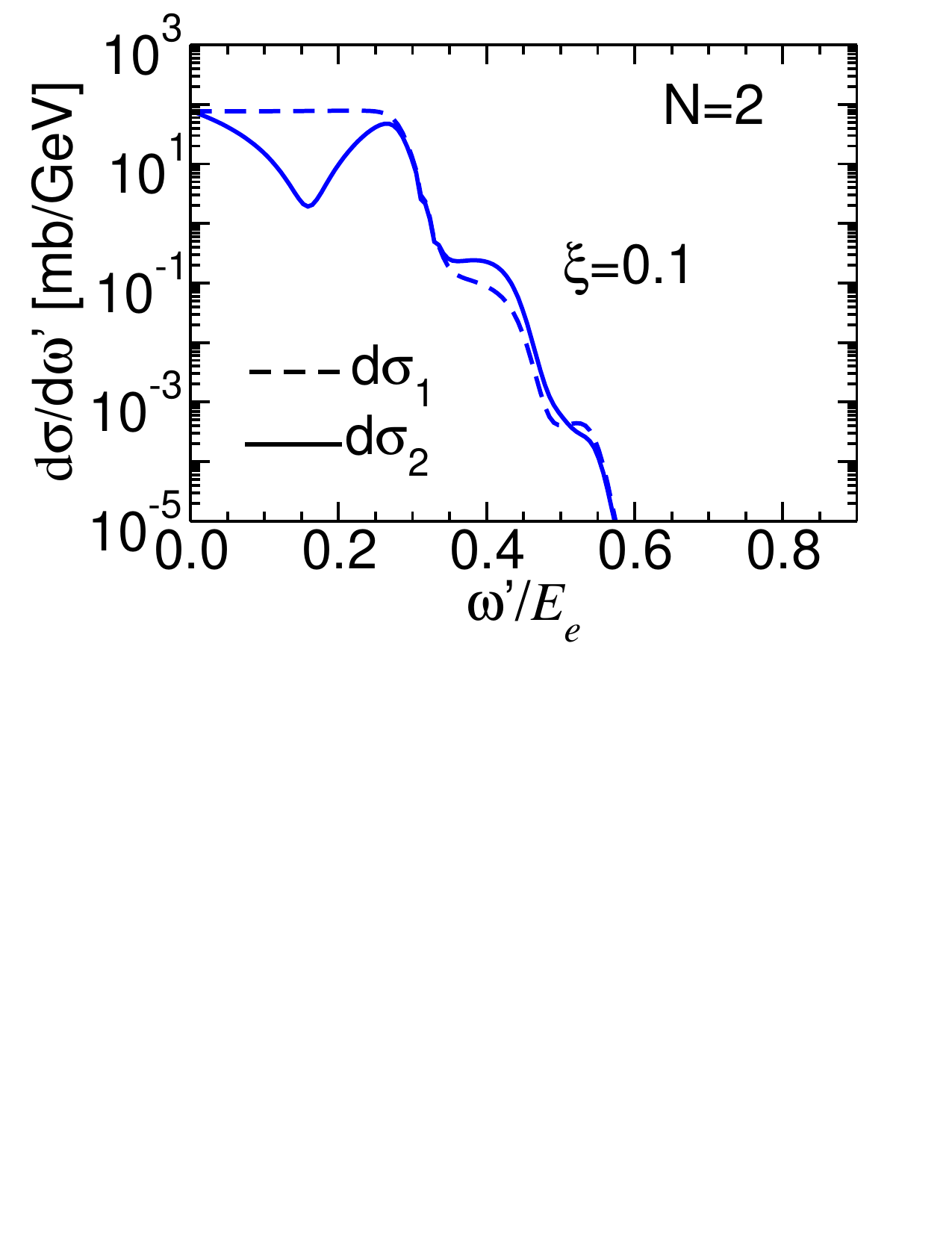}
  \includegraphics[width=0.49\columnwidth]{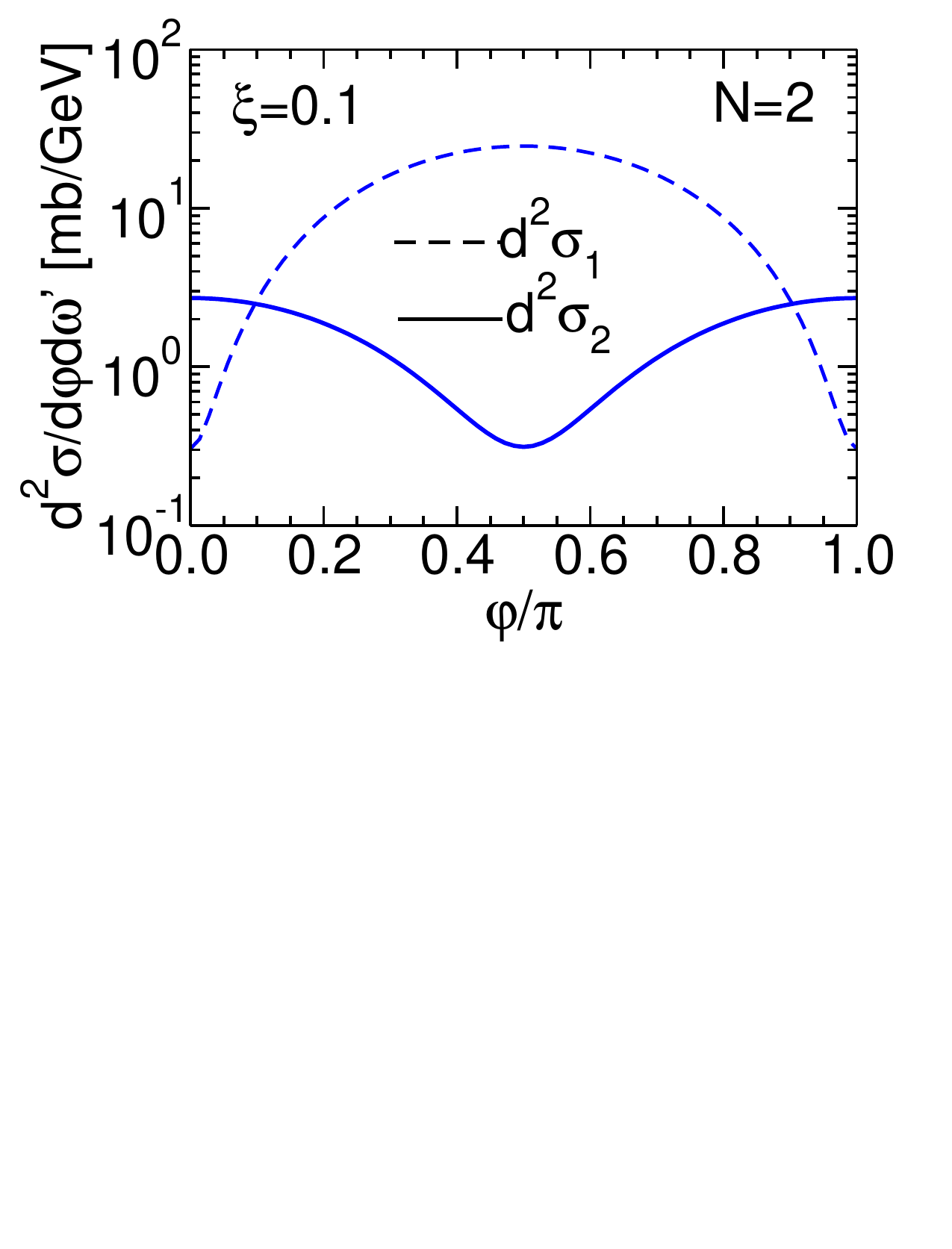} \\  
  \vspace{-2.8cm}    
  \includegraphics[width=0.49\columnwidth]{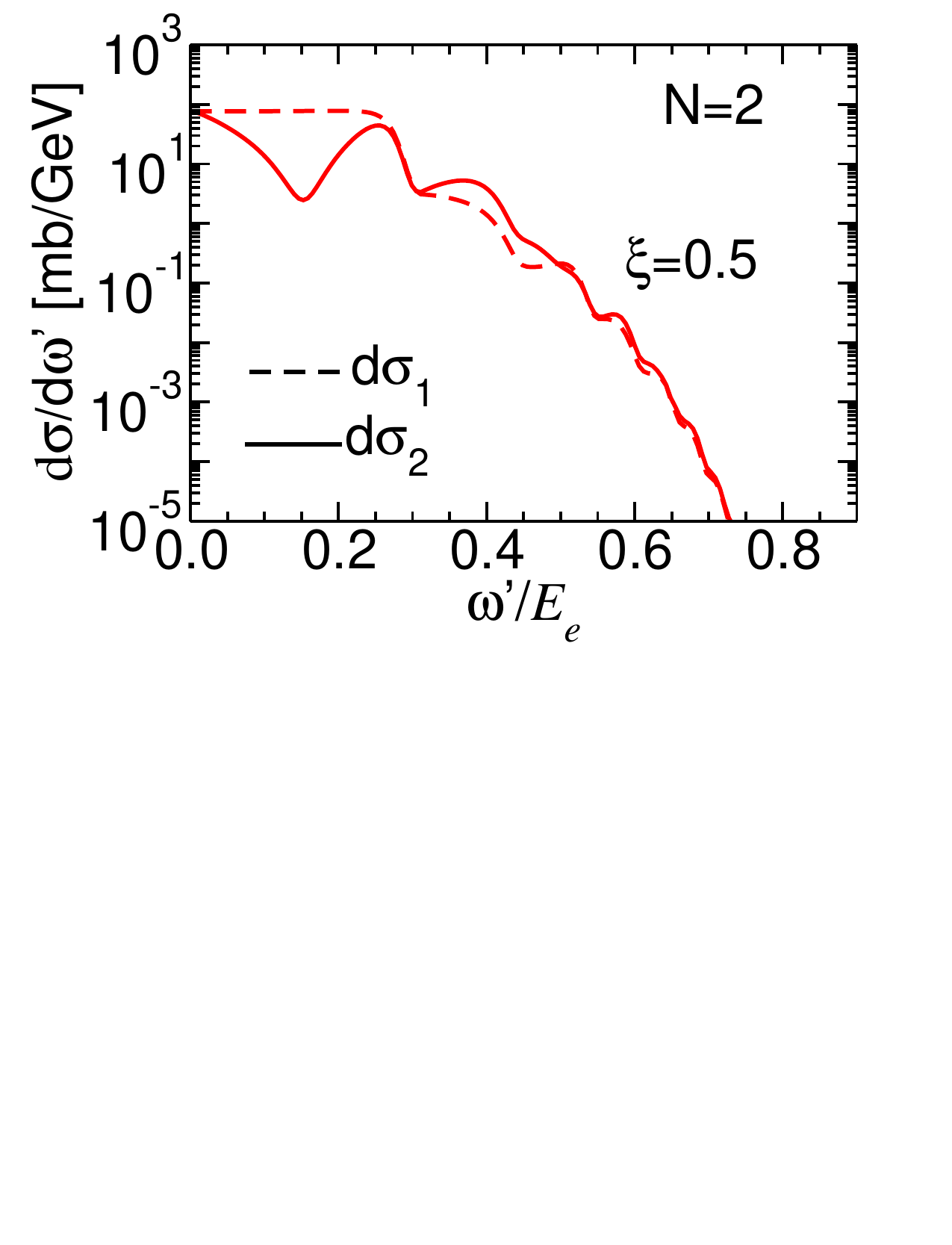}
  \includegraphics[width=0.49\columnwidth]{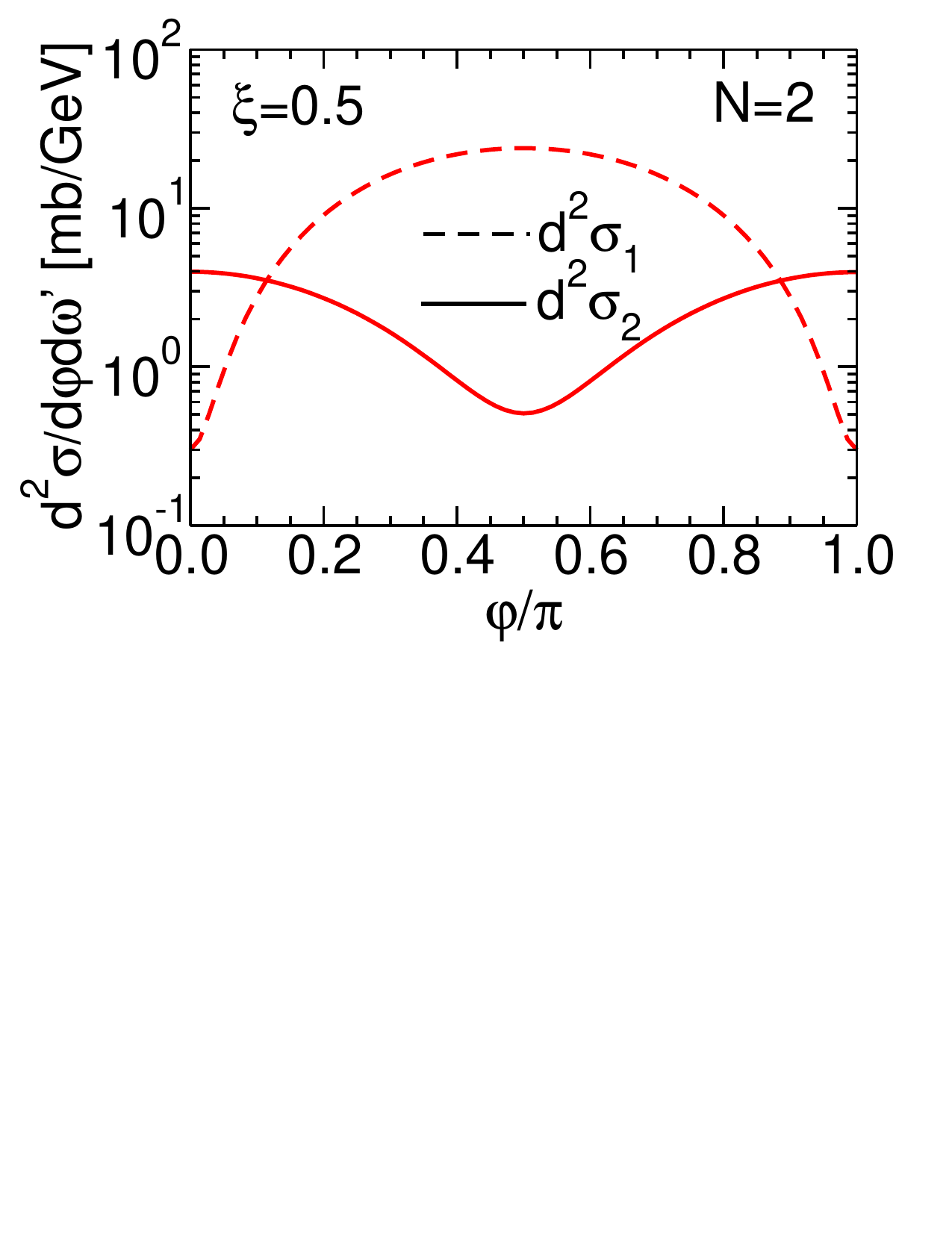} \\ 
  \vspace{-2.8cm}    
  \includegraphics[width=0.49\columnwidth]{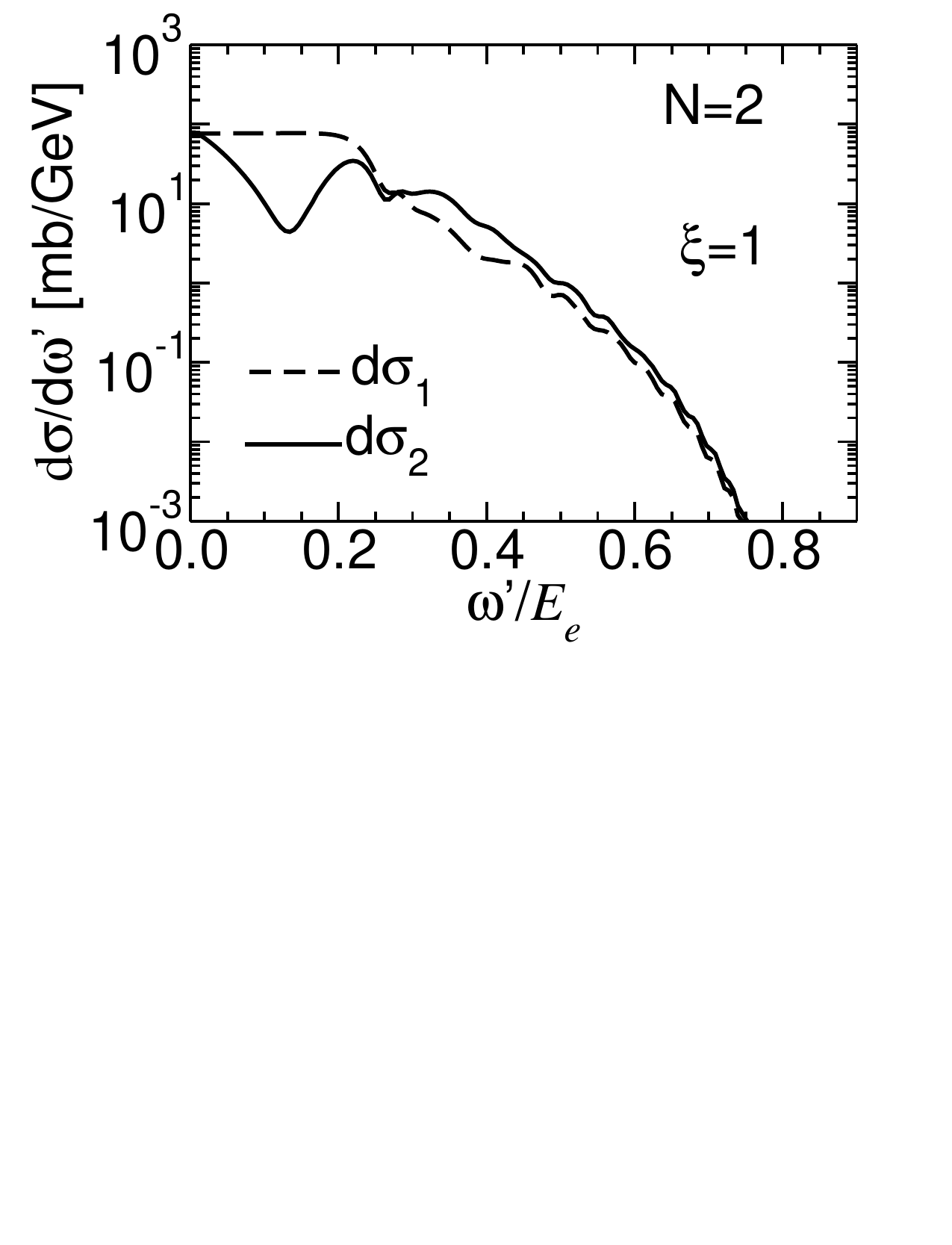}
  \includegraphics[width=0.49\columnwidth]{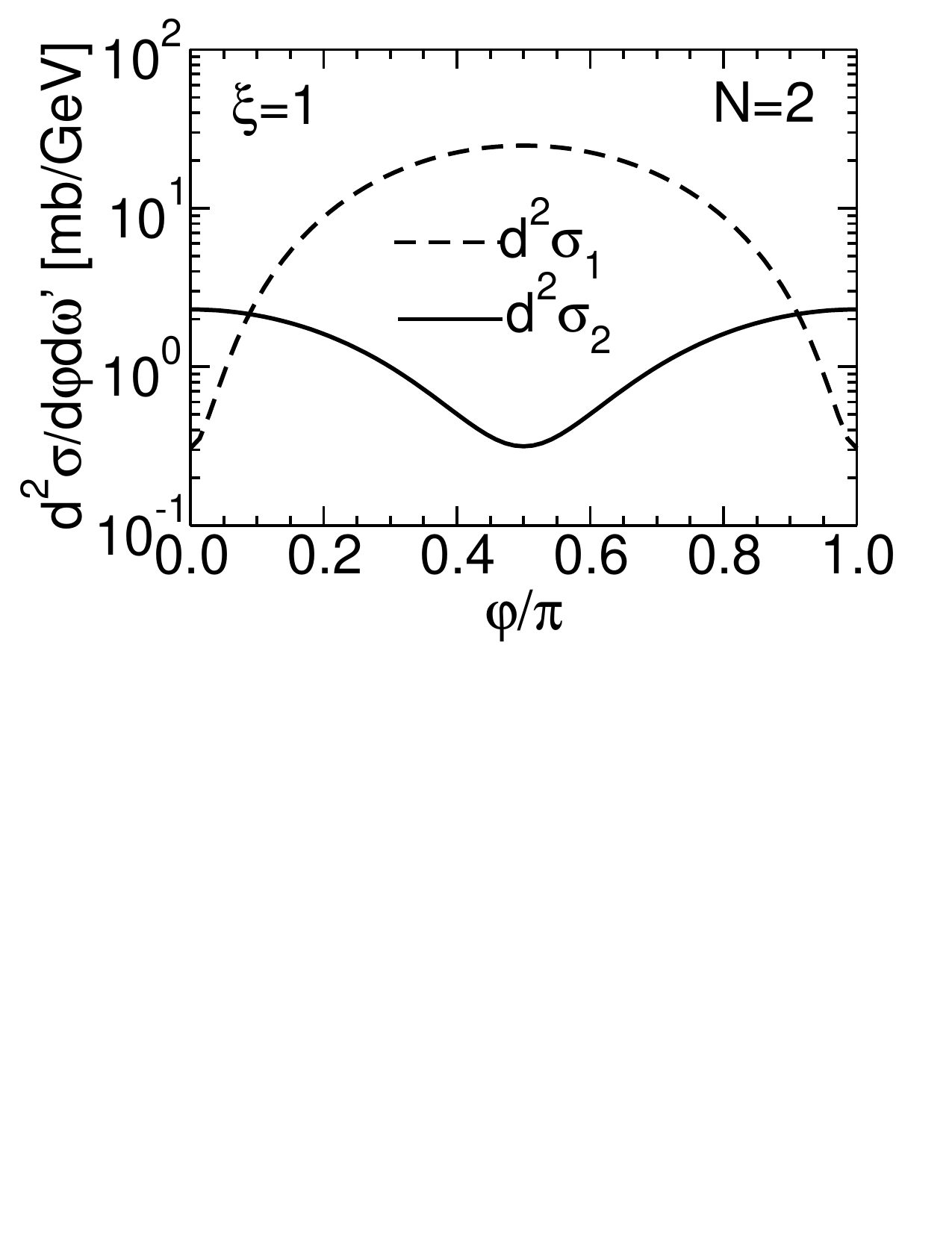}\\
       \vspace{-3.0cm}
 \caption{\small{(Color online) 
     The left panels: The differential partial cross sections
     $d\sigma_i/d\omega'$ for a number of oscillations
     in a pulse $N=2$ and different $\xi$.
     The right panels: The differential  cross sections
     $d^2\sigma_i/d\varphi d\omega'$ as functions
     of the azimuthal angle $\varphi$ at fixed $\omega'=0.2\,E_e$.
   \label{Fig:6}}}
  \end{figure}
  
  \begin{figure}[h]
  \includegraphics[width=0.49\columnwidth]{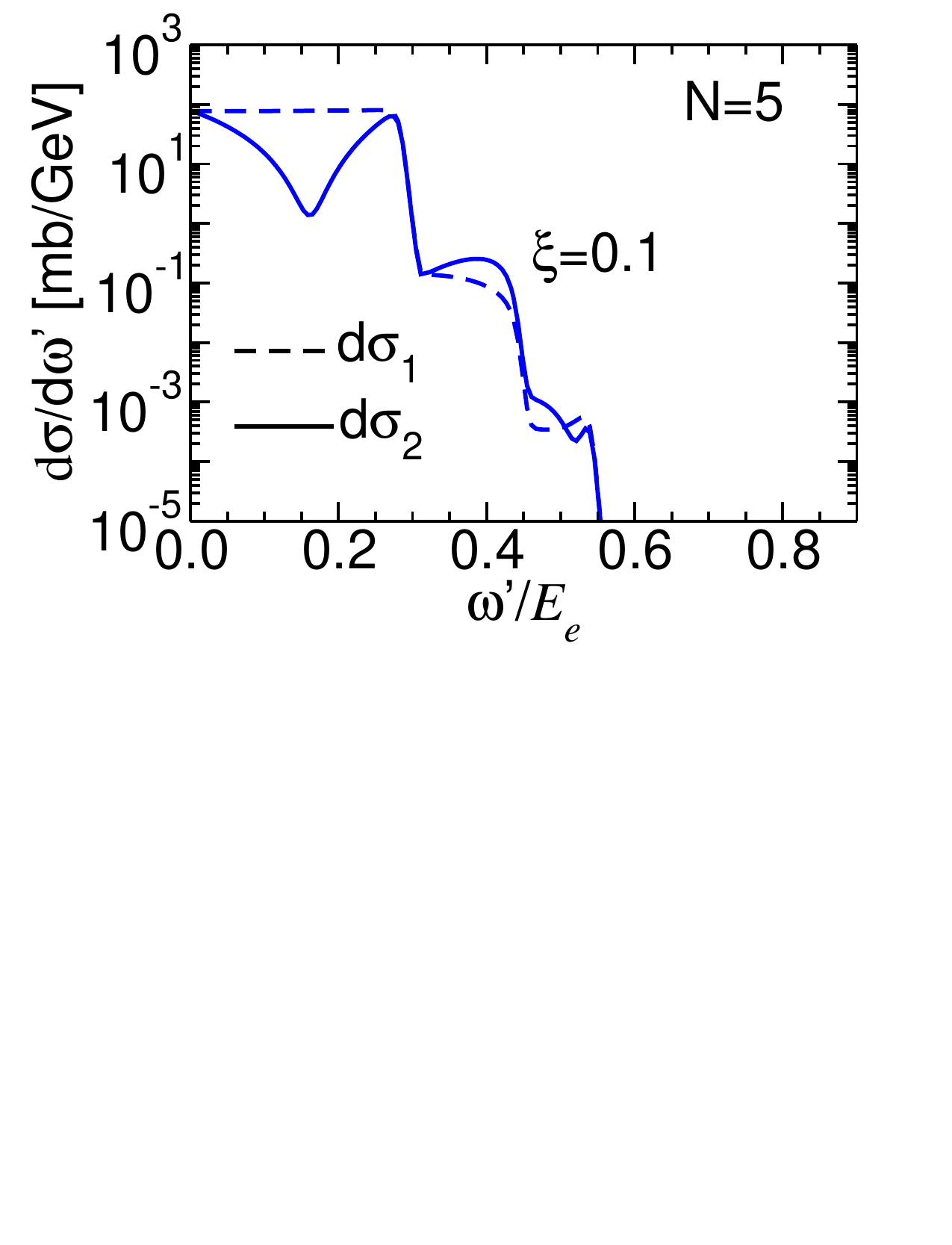}
  \includegraphics[width=0.49\columnwidth]{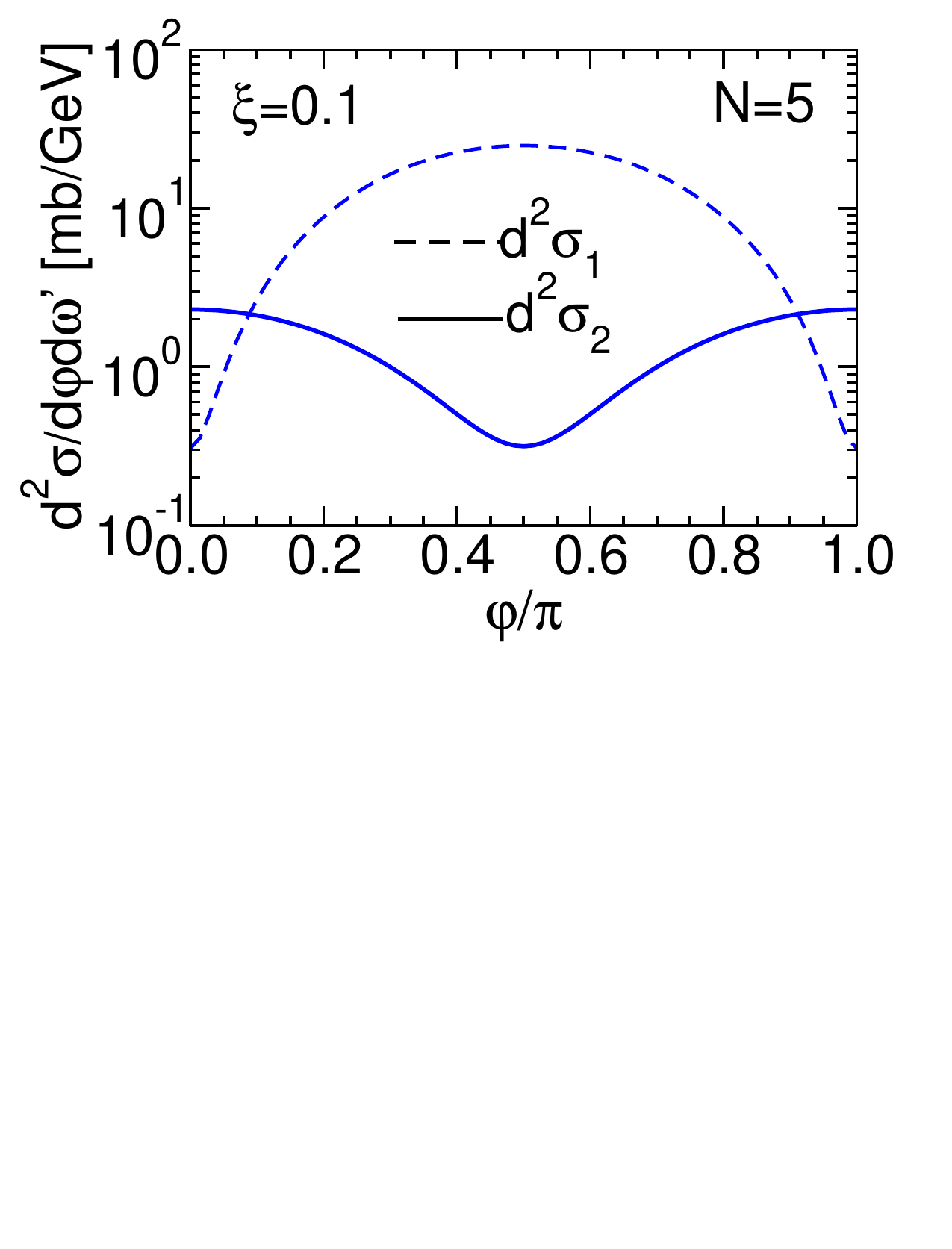} \\  
  \vspace{-2.8cm}    
  \includegraphics[width=0.49\columnwidth]{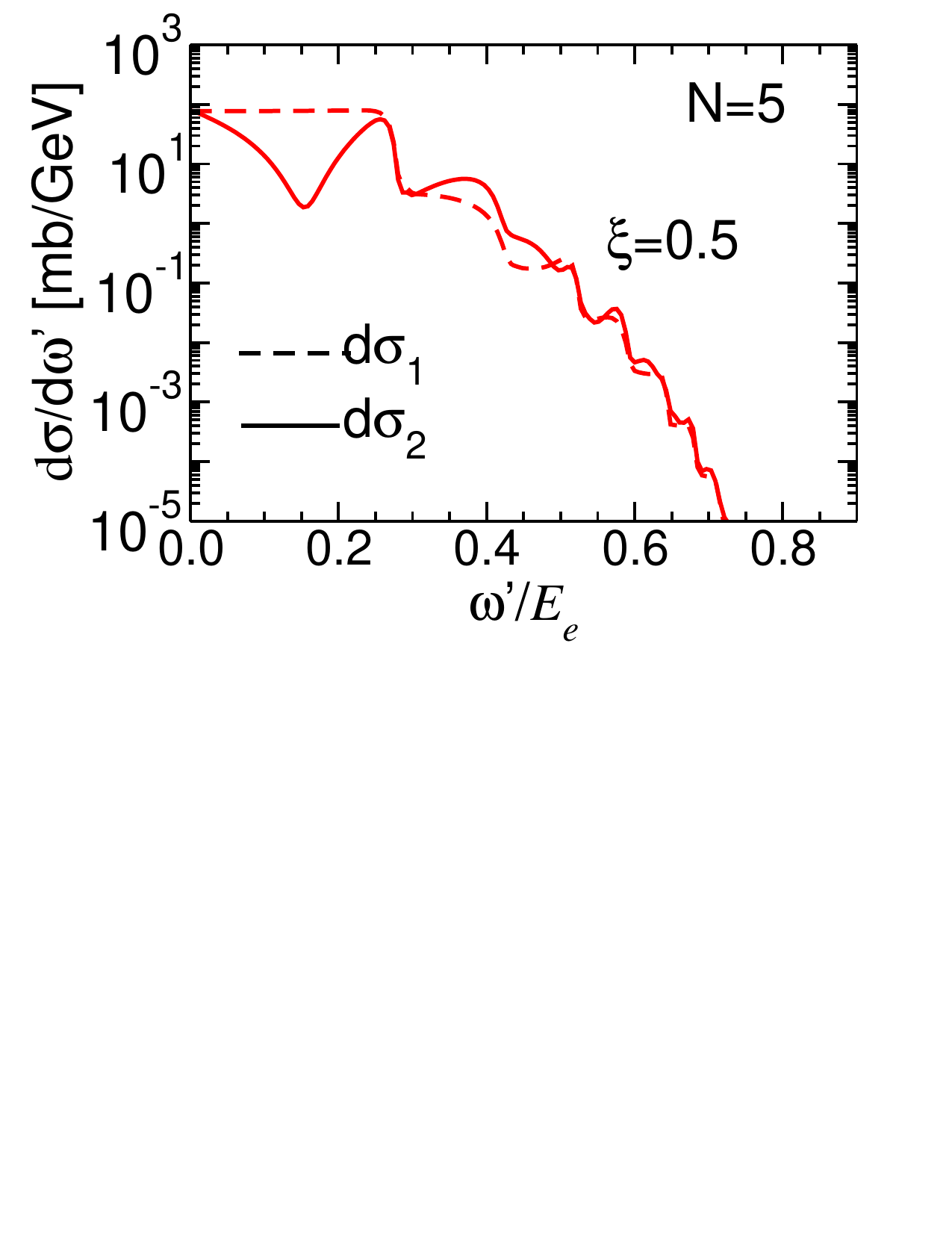}
  \includegraphics[width=0.49\columnwidth]{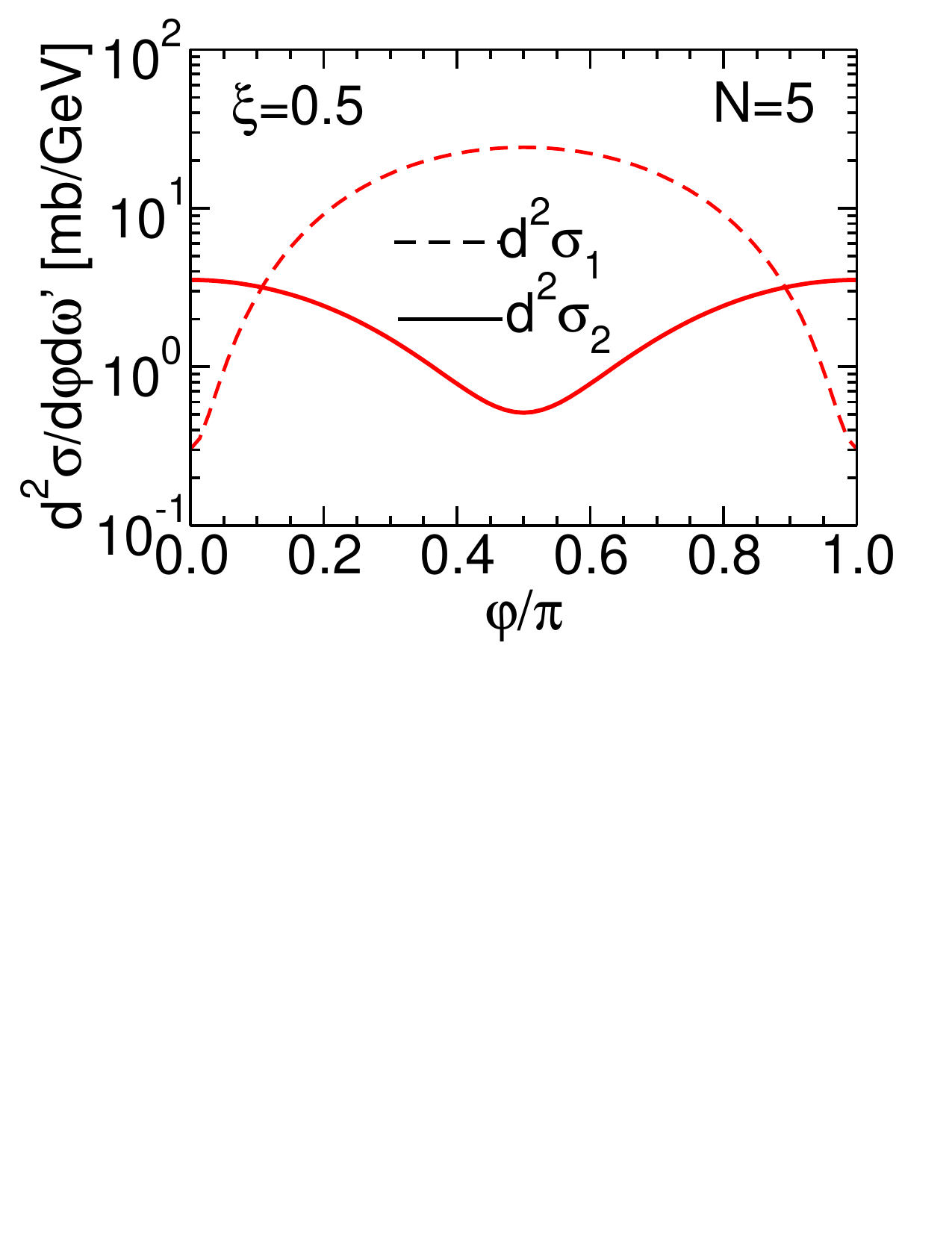} \\ 
  \vspace{-2.8cm}    
  \includegraphics[width=0.49\columnwidth]{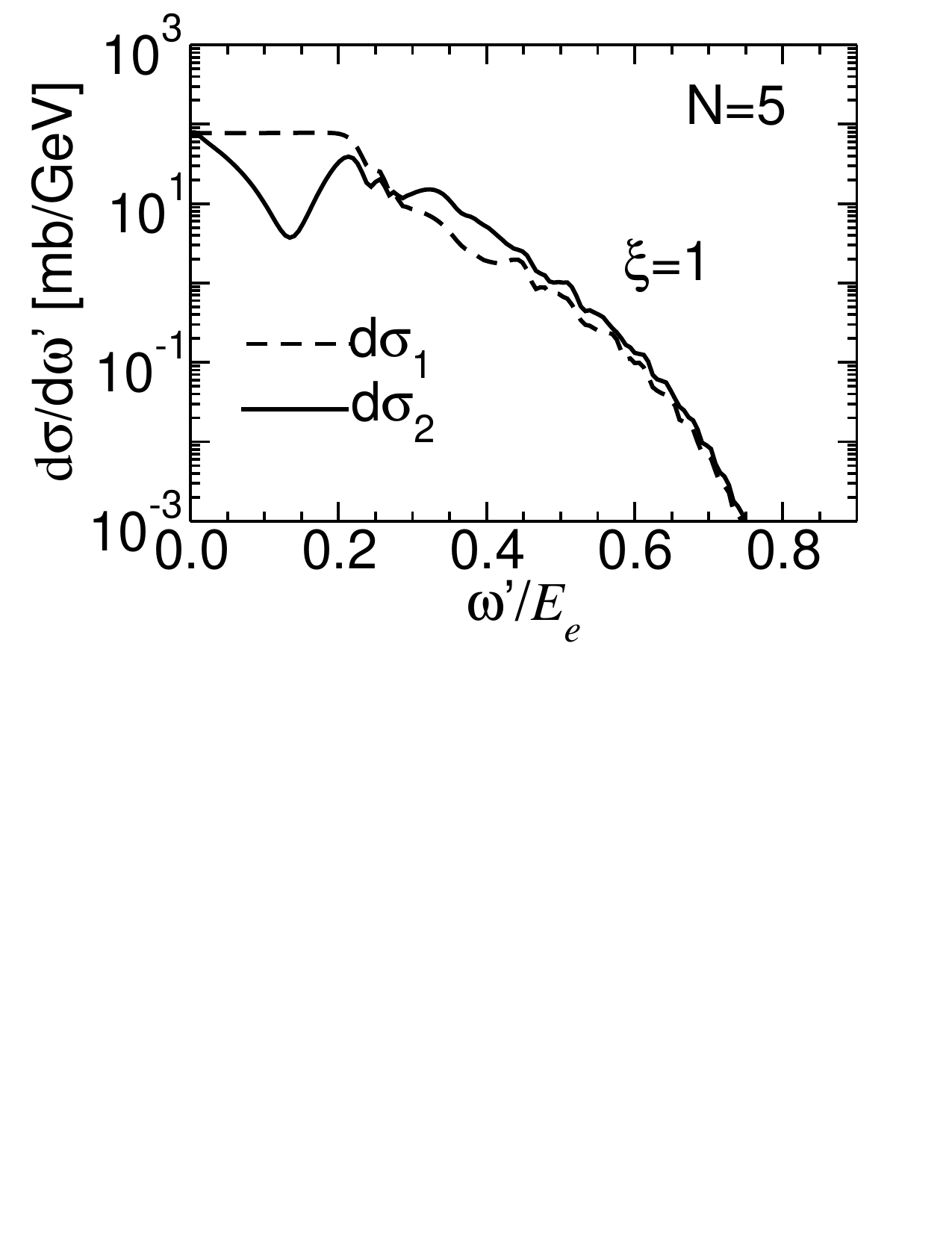}
  \includegraphics[width=0.49\columnwidth]{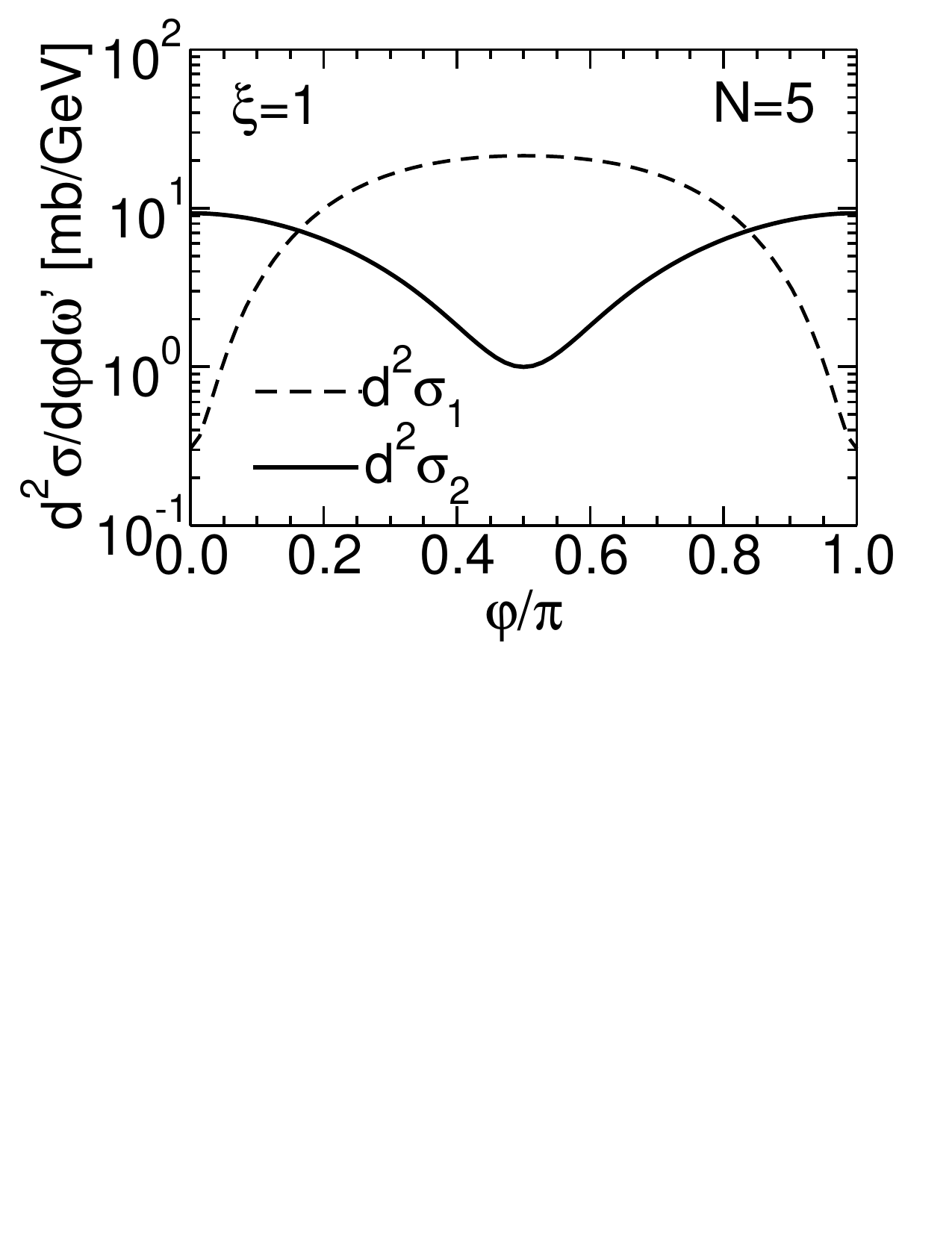}\\
         \vspace{-3.0cm}
       \caption{\small{(Color online)
       The same as in Fig.~\ref{Fig:6} but for $N=5$.     
   \label{Fig:7}}}
 \end{figure}

  \begin{figure}[h]
    \includegraphics[width=0.49\columnwidth]{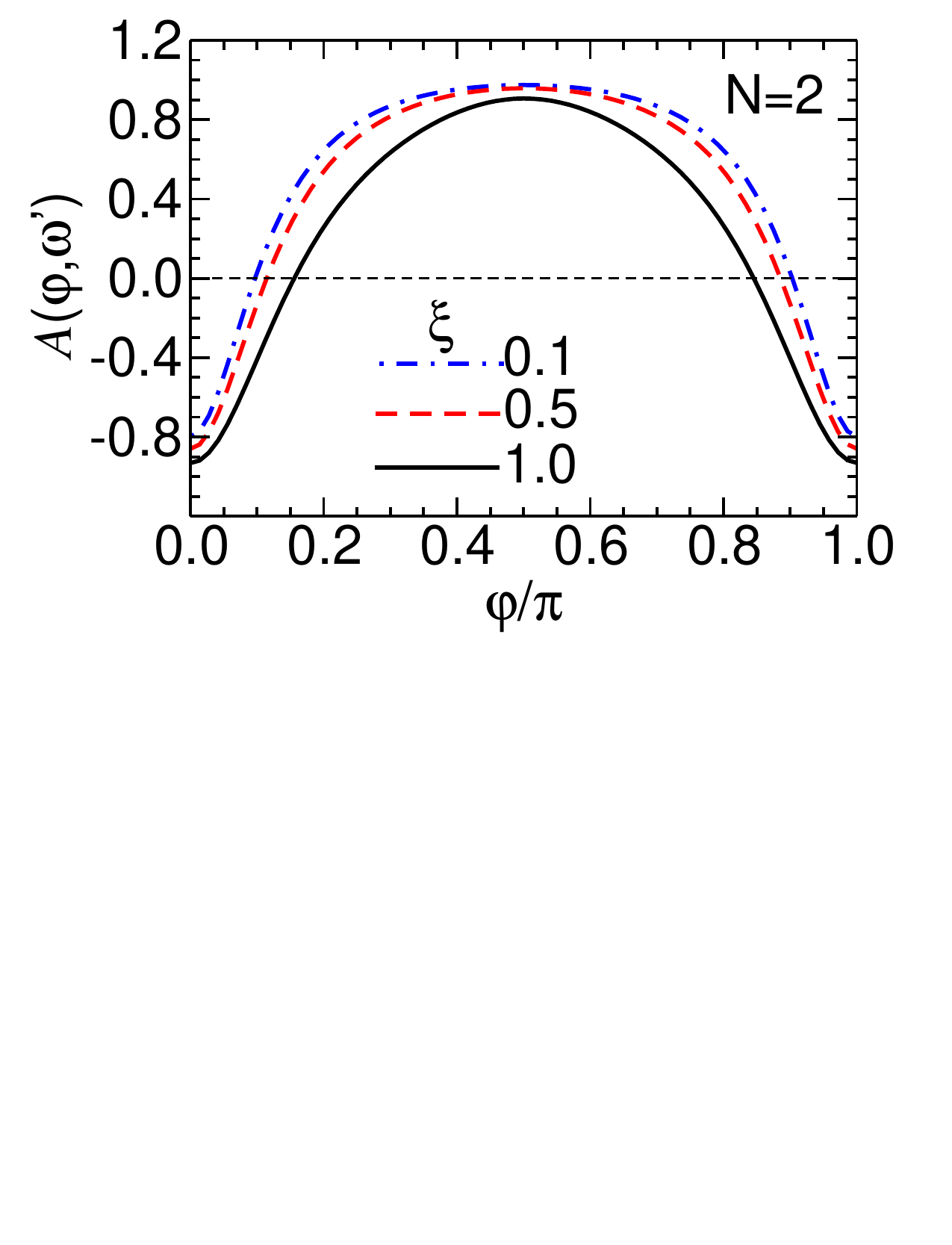}  
    \includegraphics[width=0.49\columnwidth]{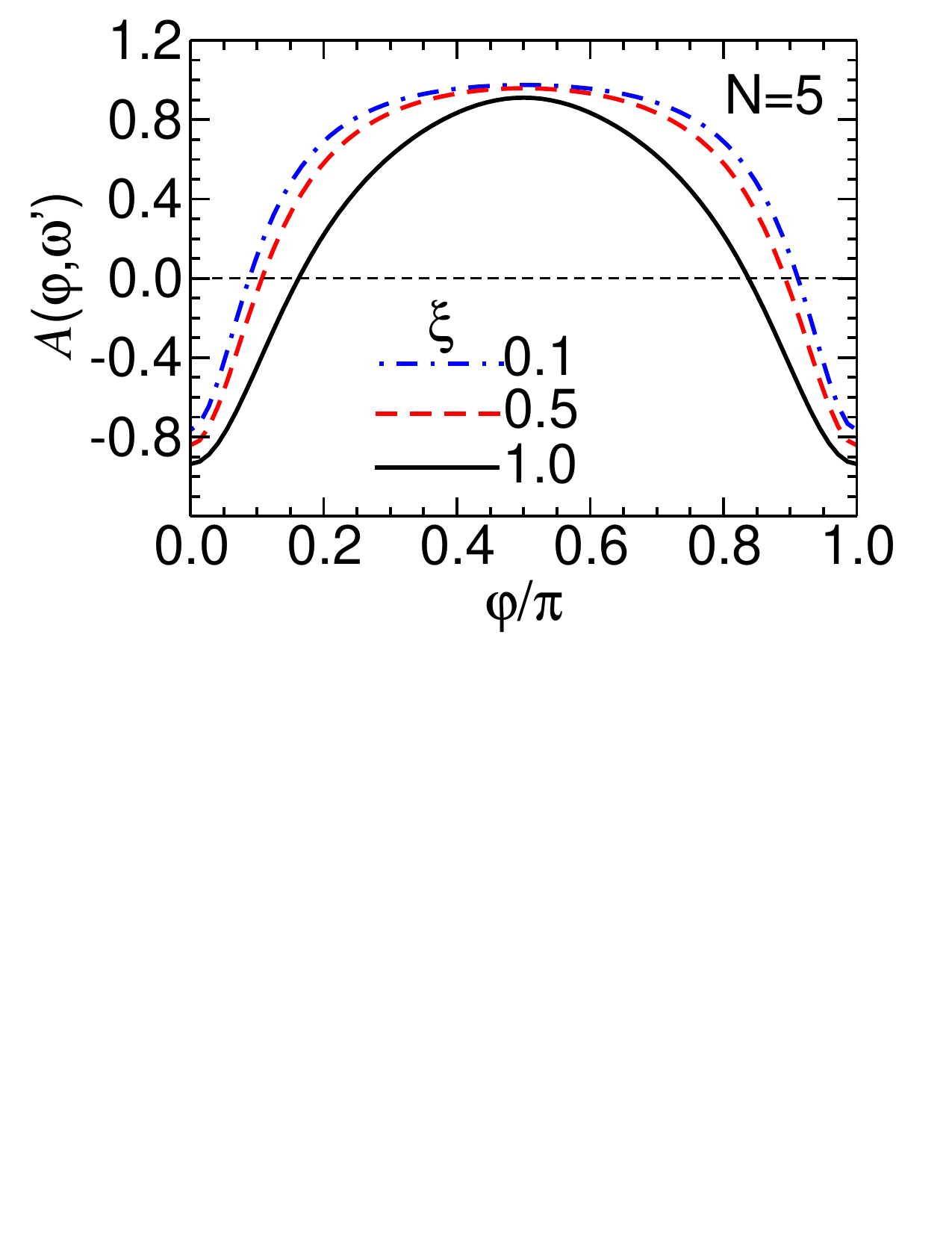}
       \vspace{-3.4cm}
 \caption{\small{(Color online) 
     The asymmetries as functions of the azimuthal angle $\varphi$
     for different $\xi$, at $\omega'=0.2\,E_e$. The left
     and right panels correspond to the number of oscillations
     in the pulse $N=2$ and 5, respectively.
   \label{Fig:8}}}
 \end{figure}

\begin{figure}[h]  
  \includegraphics[width=0.49\columnwidth]{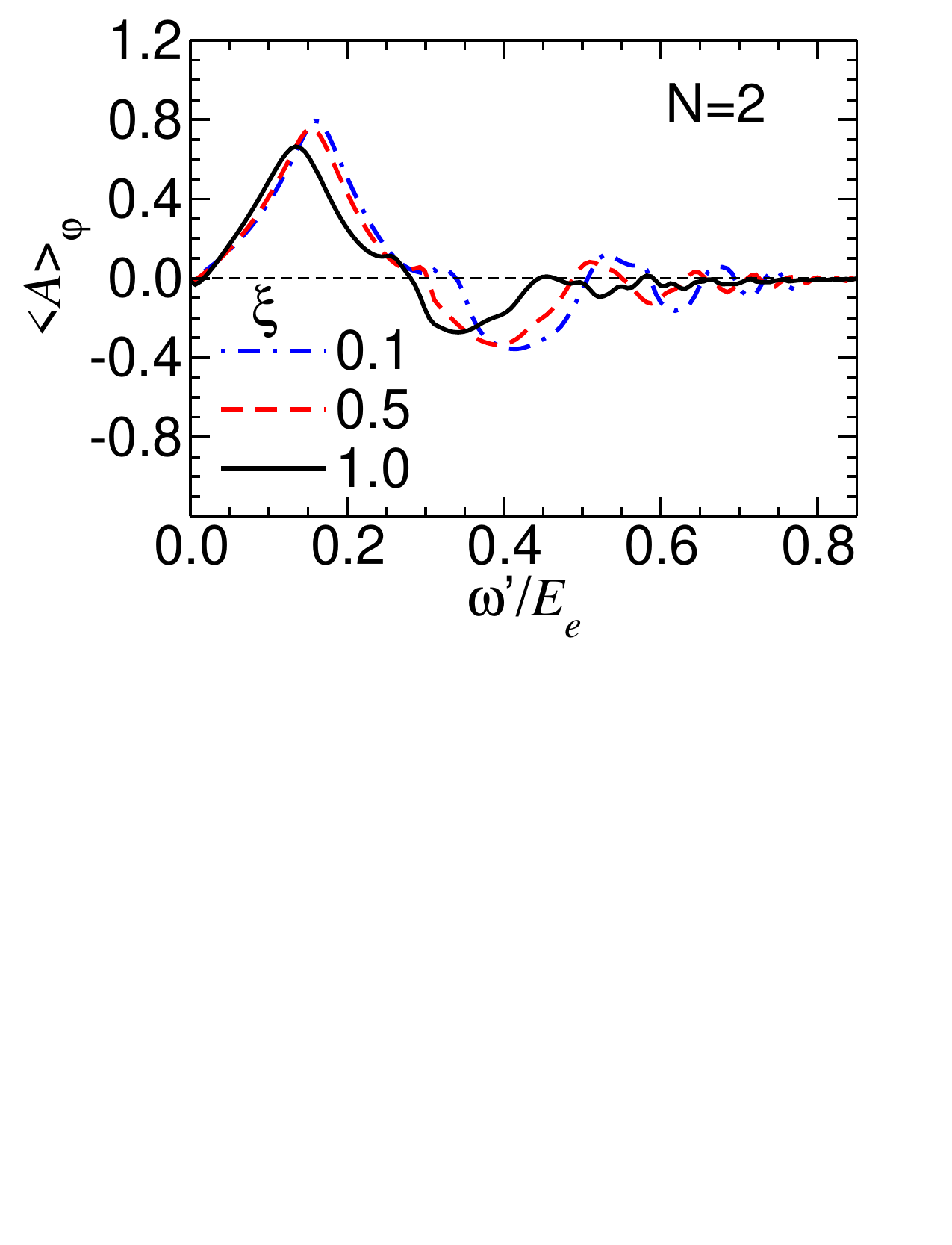}  
  \includegraphics[width=0.49\columnwidth]{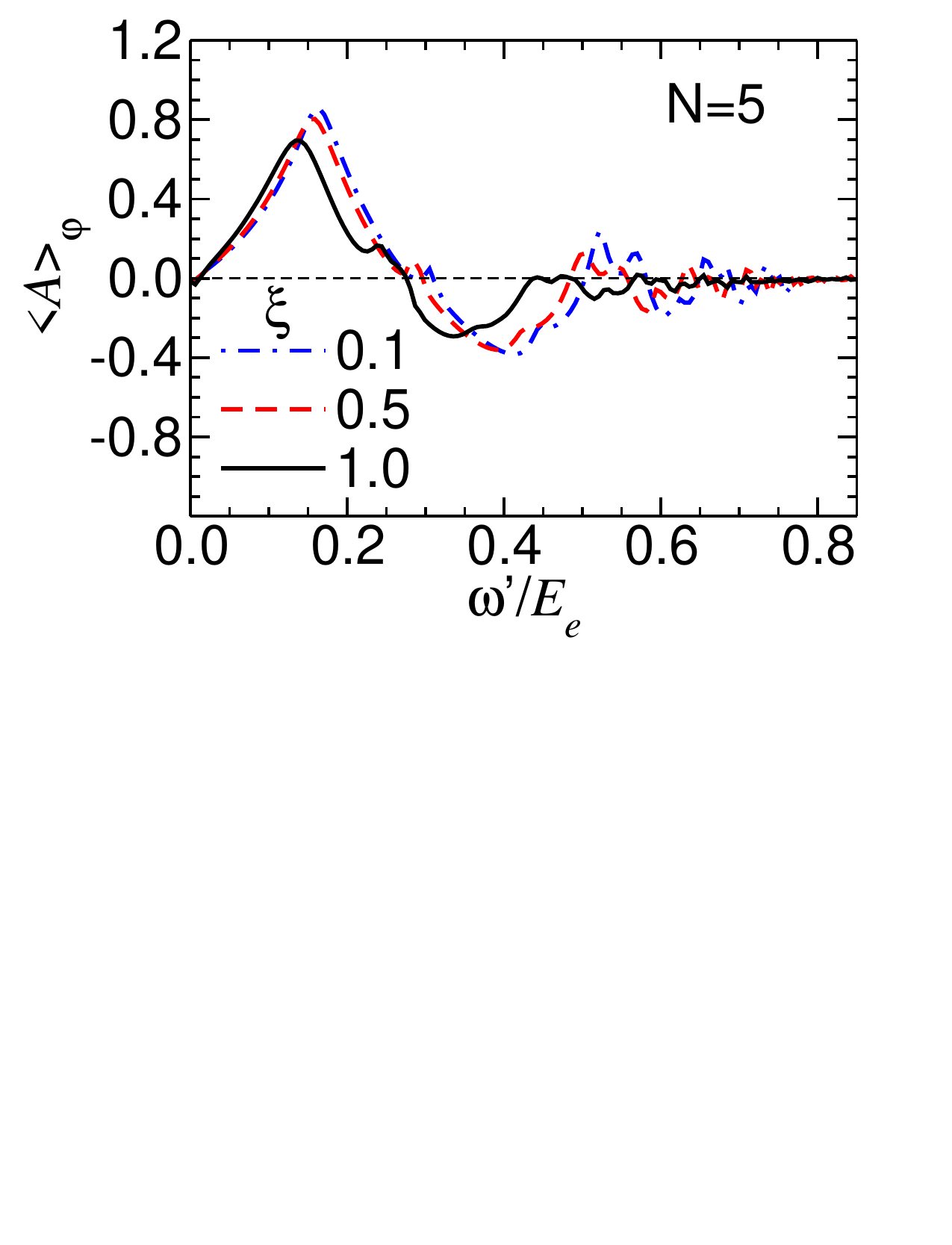}  
 \vspace*{-3.4cm}   
 \caption{\small{(Color online) 
     The average asymmetries
     $\langle{\cal A}\rangle_\varphi$ as  functions of $\omega'$
     for different $\xi$.
     The left and right panels correspond to the number of oscillations
     in the pulse $N=2$ and 5, respectively.
  \label{Fig:9}}}
\end{figure}

{In our consideration, the $S$-matrix element is determined
in the same way as in the monochromatic case, using the exact Volkov
solution for electron wave functions with e.m. four-potential~(\ref{Four_A}).
The transition matrix element is calculated using the Fourier
integral~\cite{Titov2020,Titov2014}. As a result, an auxiliary continuous variable
$\ell$ appears. The product $\omega\ell$ has the meaning of the fraction
of the laser beam energy involved in the non-linear Compton process.
Together with the replacement of summation over discrete
$l$ by the integral over continuous $d\ell$ in the transition matrix,
there is one more difference:
now the pre-exponential factor and the exponential factor
in electron wave functions containing linear and quadratic field terms
of $A(\phi)$ form new basis functions ${\widetilde A_m}$.
The structure of the matrix element
remains similar to the monochromatic case~(\ref{II002}).
The new basis functions ${\widetilde A_m}$ read~\cite{impact}}
 \begin{eqnarray}
 \widetilde A_m(\ell)
 =\frac{1}{2\pi}\int\limits_{-\infty}^{\infty}d\phi\,f^m(\phi)
 \cos^m(\phi)\,{\rm e}^{i\ell\phi
 -i{\cal P}(\phi)}~,
 \label{III4}
 \end{eqnarray}
where
\begin{eqnarray}
&&{\cal P}(\phi) =
\tilde\alpha(\phi)- \tilde\beta(\phi)~,\label{II5}\\
&&\tilde\alpha(\phi)=\alpha'\,h_1(\phi),
\qquad\tilde\beta(\phi)=4\beta\,h_2(\phi)\label{III6}	
\end{eqnarray}
with
\begin{eqnarray}
h_n(\phi)= \int\limits_{-\infty}^{\phi}d\phi'
f^n(\phi')\cos^n\phi'~,\label{III7}
\end{eqnarray}
where $\alpha'$ and $\beta$ are determined in (\ref{ab6})
with
$z={2\xi\ell}\sqrt{{u/u_\ell}(1-u/u_\ell)}$
and $u_\ell=2\ell\chi/\xi$.\\
The function $\widetilde A_0(\ell)$ is regularized using the
prescription of \cite{Boca-2009} which leads to the identity
\begin{eqnarray}
  {\alpha'}\widetilde A_1(\ell)=\ell\widetilde A_0(\ell)
  + 4\beta\widetilde A_2(\ell)~.
\label{III8}
\end{eqnarray}

The differential cross sections
$d^2\sigma_i$ have the form
of equations~(\ref{II6_},\ref{II7}) with the replacement
of $\sum\limits_{\ell}$ by $\int\limits_{\ell_{\rm min}}^{\infty}d\ell$,
where the lower limit of the integral is $\ell_{\rm min}=um^2/2k\cdot p$ and
the replacement  $A_m\to\widetilde A_m$
\begin{eqnarray}
{d^2 \sigma_{1}}
&=&\frac{2\alpha^2d\varphi d\omega'}{\xi\chi m^2E_e~N_0}
\int\limits_{\ell_{\rm min}}^{\infty}~d\ell
\, [\xi^2~\vert\widetilde{A}_1\vert^2\sin^2\varphi\nonumber\\
  &+&\xi^2\frac{u^2}{4(1+u)}\left(\vert\widetilde{A}_1\vert^2
  -{\rm Re}\widetilde{A}_0\widetilde{A}^*_2\right)]~,
\label{III1}\\
{d^2 \sigma_{2}}
&=&\frac{2\alpha^2 d\varphi d\omega'}{\xi\chi m^2E_e~N_0}
\int\limits_{\ell_{\rm min}}^{\infty}~d\ell
\, [-\vert\widetilde{A}_0\vert^2
  -\xi^2\,\vert\widetilde{A}_1\vert^2\sin^2\varphi\nonumber\\
  &+&\xi^2 \left(1+\frac{u^2}{4(1+u)}\left(\vert\widetilde{A}_1\vert^2
  -{\rm Re}\widetilde{A}_0\widetilde{A}^*_2\right)\right)]~,
\label{III2}
\end{eqnarray}
where replacement $|\vec q-l\omega|\simeq |\vec p|\simeq E_e$ is used.
The unpolarized cross section is the sum
\begin{eqnarray}
  \label{III22}
  {d^2 \sigma} = {d^2 \sigma_{1}} + {d^2 \sigma_{2}}~.
\end{eqnarray}

{Utilization of basis functions $\widetilde{A}_{1,2}$ in (\ref{III4})
lead to a multidimensional integral for $d\sigma_{1,2}/d\omega'$ which needs
large computing resources. We calculate them as follows.
For given wave envelope functions $f(\phi)$,
we first calculate two  arrays of auxiliary functions
$h_{1,2}(\phi)$~(\ref{III7})
in the range $\phi_{\rm  min}\le\phi\le\phi_{\rm max}$,
where $\phi_{\rm max, min}=\pm\,30\,\Delta$
with $10 ^5$ points ($\Delta=\pi N$ and $N$ is the number
of oscillations in a pulse).
In fact, $\phi_{\rm max, min}$ are the limits of integration in (\ref{III4}).
Then, using the FORTRAN interpolation software PARINT, we extract
the phase function ${\cal P}(\phi)$ (\ref{II5})
and determine the complex integrand in
(\ref{III4}) at any point in the integration
domains in (\ref{III1}), (\ref{III2}).
This technique reduces the dimension of the integrals for $d\sigma_i$
and the time of their calculation.}
The function $\widetilde{A}_{0}$ is determined by Eq.(\ref{III8}).

The flux parameter $N_0$ in (\ref{III1},\,\ref{III2}) is related
to the average square of the e.m. strength and expressed thought
the envelope function as integral
$1/2\pi\int_{-\infty}^{\infty}d\phi(f^2(\phi) + {f'}^2(\phi))\cos^2\phi$
and for chosen $f(\phi)$ is equal
$N_0={\Delta}/{2\pi}(1 +1/{3\Delta^2})$.

Below we will limit ourselves to the number of oscillations
in a pulse $N=2$ and 5 and field strength $\xi=0.1$, 0.5 and 1.
The unpolarized cross sections for a finite pulse are exhibited
in Fig.~\ref{Fig:5}. This result is close to our previous works
(see, for example ~\cite{TitovPEPAN,impact}).
For completeness and easy references
we also exhibited in Fig.~\ref{Fig:5} result for monochromatic
background shown by
solid thin curves. Again, one can see a step-like behavior of the
cross section at small $\xi$.  The result for N=5 is closer to the
monochromatic background and, for N=10 
they are almost identical~\cite{impact}.

The partial cross sections $d\sigma_i/d\omega'$
for N=2 and 5 are shown in the left panels
of Figs.~\ref{Fig:6} and \ref{Fig:7}, respectively.
In general, the results for $N=2$ and 5 are close to each other.
It can be seen that the result for the finite pulse background
is qualitatively close to the result for the monochromatic PW background shown
in Figs~\ref{Fig:2} and 3: now the cross sections are
slightly smoother functions of $\omega'$ .

The same is true for the azimuthal angle distributions of differential
cross sections shown in the right panels of Figs.~\ref{Fig:6} and
\ref{Fig:7}, and asymmetries shown in Fig.~\ref{Fig:8}.
The average asymmetries  shown in~Fig.~\ref{Fig:9}
resemble the result of the monochromatic PW calculation, cf. Fig.~\ref{Fig:4} (left).
The symmetry property (\ref{symmetry}) is also valid for a finite pulse,
which is confirmed by direct numerical calculation.

To summarize this part, we can conclude that in general,
our result for a finite pulse is similar to the case of the
monochromatic plane wave background, i.e.
in coplanar geometry with $\varphi=0$ and in a frame with $\varphi=\pi/2$
the recoil photon $\gamma'$ is polarized almost
{\it parallel to the polarization of the laser beam} with a degree
polarization ${\cal P}$ close to unity. 

\section{Ultra-large pulse intensity, ${\mathbf\xi}\gg1$}
For large values of $\xi\gg1$ the main contribution to
probabilities of non-linear processes
comes from the central part of the wave envelope and the diagonal terms
in the amplitude squared with $\ell=\ell'$~\cite{TitovPRA2013}.
Therefore, as in~\cite{TK2020}, for further analysis we use an extension
of the formalism developed by Nikishov and Ritus~\cite{RitusGroup},
with explicit consideration of the $\gamma'$ polarization.
Strictly speaking, the validity of this approximation
for the chosen variables depends on the value of the recoil
photon energy $\omega'$. We will discuss this issue later.

The partial and unpolarized
differential sections $d^2\sigma_i$, $d^2\sigma$
have the following form:
\begin{eqnarray}
{d^2\sigma_{1}}
&=&
\frac{4\alpha^2 d\varphi d\omega'}{m^2\xi\chi\,E_e}
\int\limits_{\ell_{\rm min}}^\infty d\ell
   \, [\xi^2~\widehat{A}_1^2\sin^2\varphi\nonumber\\
  &+&\xi^2\frac{u^2}{4(1+u)}\left(\widehat{A}_1^2
  -\widehat{A_0}\widehat{A}_2\right)]~,
\label{IV1}\\
{d^2\sigma_{2}}
&=&
\frac{4\alpha^2 d\varphi d\omega'}{m^2\xi\chi\,E_e}
\int\limits_{\ell_{\rm min}}^\infty d\ell
\, [-\widehat{A}_0^2 -\xi^2\,\widehat{A}_1^2\sin^2\varphi\nonumber\\
  &+&\xi^2 \left(1+\frac{u^2}{4(1+u)}
  \left(\widehat{A}_1^2-\widehat{A}_0\widehat{A}_2\right)\right)]~.
\label{IV2}\\
{d^2\sigma}
&=&{d^2\sigma_{1}}
+{d^2\sigma_{2}}~,
\label{IV12}
\end{eqnarray}
where $\ell_{\rm min}= u\xi(1+\xi^2/2)/2\chi$.


The bi-linear combinations of $\widehat{A}_k$ are expressed
through the Airy functions $\Phi$ and their derivatives $\Phi'$
\begin{eqnarray}
  \widehat{A}_0^2&=&\frac{g^2}{2\pi^2}\Phi^2(y)~,\qquad
  g^2=\frac{4}{\xi^2\sin^2\psi}\frac{\sigma}{y},\nonumber\\
  \widehat{A}_1^2&=&\frac{g^2}{2\pi^2}(\rho^2\Phi^2(y)
  +\frac{\sigma}{\xi^2y}{\Phi'}^2(y))~,\nonumber\\
  \widehat{A}_0\widehat{A}_2
  &=&\frac{g^2}{2\pi^2}(\rho^2-\frac{\sigma}{\xi^2})\Phi^2(y)~,
\label{IV3}  
\end{eqnarray}
where the argument of the Airy function $y$ is related to
the auxiliary variables  $\rho=\cos\psi$, $\tau$ and $\sigma$ as
\begin{eqnarray}
  y&=&\left(\frac{u}{2\chi\sin\psi}\right)^{2/3}\sigma~,\quad
  \sigma=1+\tau^2\nonumber\\ 
 \rho^2&=&\frac{1}{\xi^2}(1+\frac{\xi^2}{2})(\frac{u_\ell}{u}-1)\cos^2\varphi
  \nonumber\\
  \tau^2&=&(1+\frac{\xi^2}{2})(\frac{u_\ell}{u}-1)\sin^2\varphi
\label{IV4}
\end{eqnarray}
with $u_{\ell}=2\ell\chi/\xi(1+\xi^2/2)$.

The validity of our approach at $\xi\gg1$ is determined by condition
$\ell_{\rm min}\gg1$. Assuming that for $\omega'\ll E_e$, $u\simeq \omega'/E_e$
we have 
\begin{eqnarray}
\omega'/E_e\gg\omega'_{\rm min}/E_e=4\chi/\xi^3~.   
\label{omegamin}  
\end{eqnarray}
For $E_e=16.5$~GeV and $\omega=1.55$~eV,
$\omega'_{\rm min}/E_e\simeq 0.78/\xi^2$.
Indeed, this $\omega'_{\rm min}/E_e$ is a small value, but nevertheless
it should be remembered that it is finite.
\begin{figure}[h]
  \includegraphics[width=0.49\columnwidth]{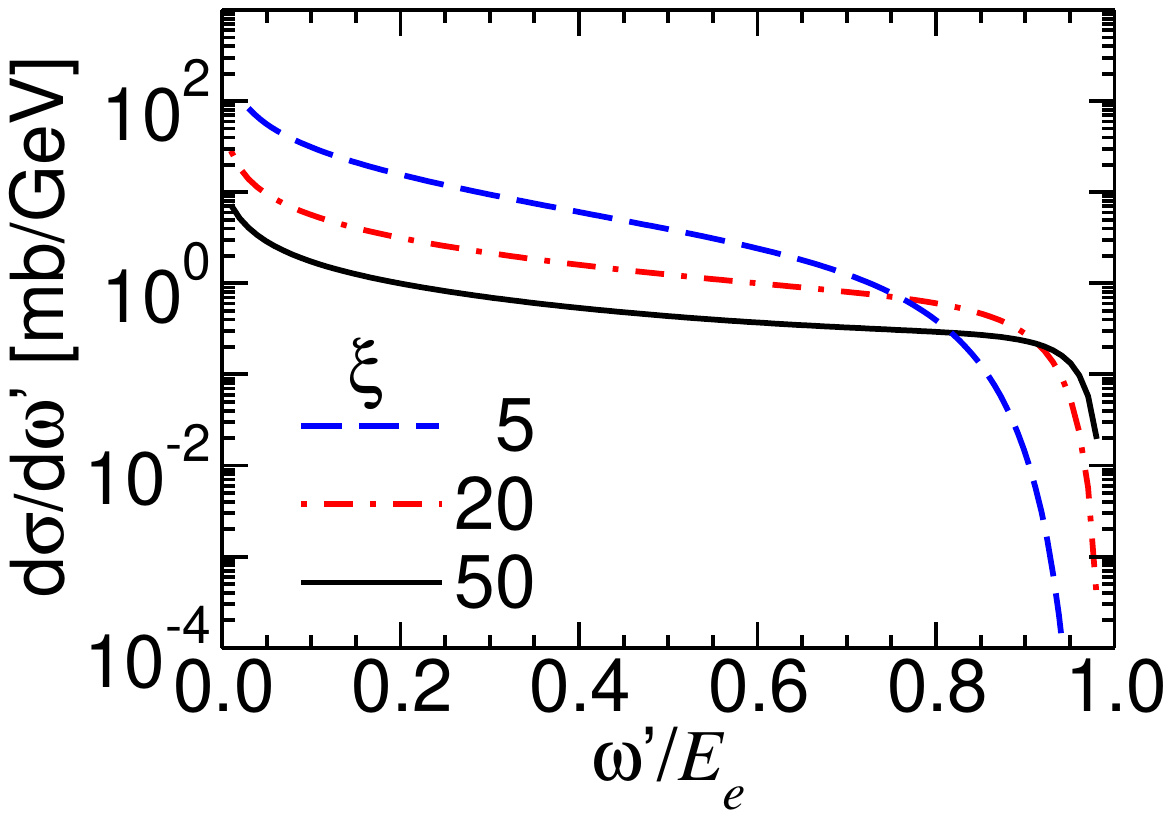}
  \includegraphics[width=0.49\columnwidth]{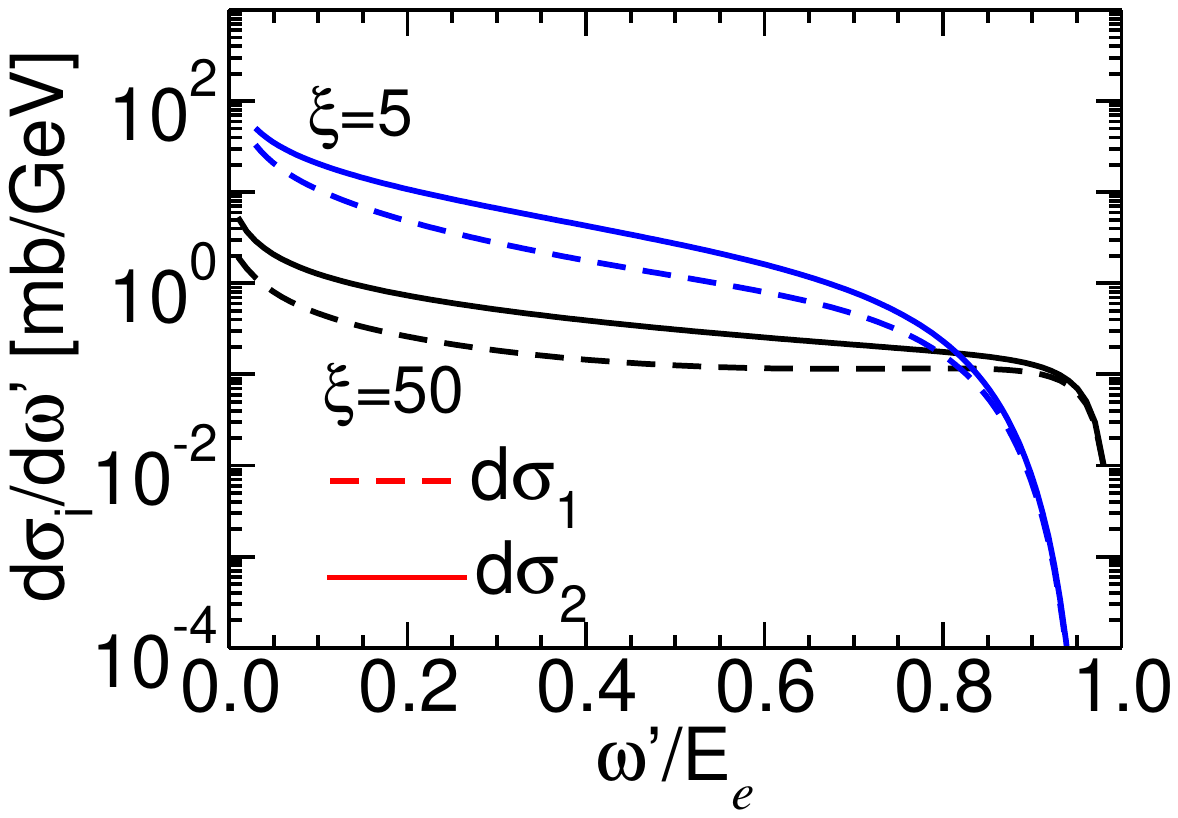}  
 \vspace{-3.4cm} 
 \caption{\small{(Color online) 
     The left panel: The unpolarized differential cross sections $d\sigma/d\omega'$ 
     as functions of $\omega'$ for $\xi=5,\ 20$ and 50
     shown by the blue dot-dashed, red dashed, and black solid
     curves, respectively.
     The right panel: 
     The partial cross sections $d\sigma_i/d\omega'$  
     as functions of $\omega'$ for $\xi=5$ and 50. 
     \label{Fig:10}
 }}
\end{figure}
The unpolarized cross sections $d\sigma/d\omega'$ integrated over
the azimuthal angle $\varphi$ for $\xi=5,$ 20 and 50 are shown in
the left panel of Fig.~\ref{Fig:10}. Here and below we consider $\omega'$ in the range
$\omega'_{\rm min}<\omega'<E_e$.
The cross sections decrease
monotonically with increasing $\omega'$, disappearing
in the limit $\omega'\to E_e$.

\begin{figure}[h]
\includegraphics[width=0.49\columnwidth]{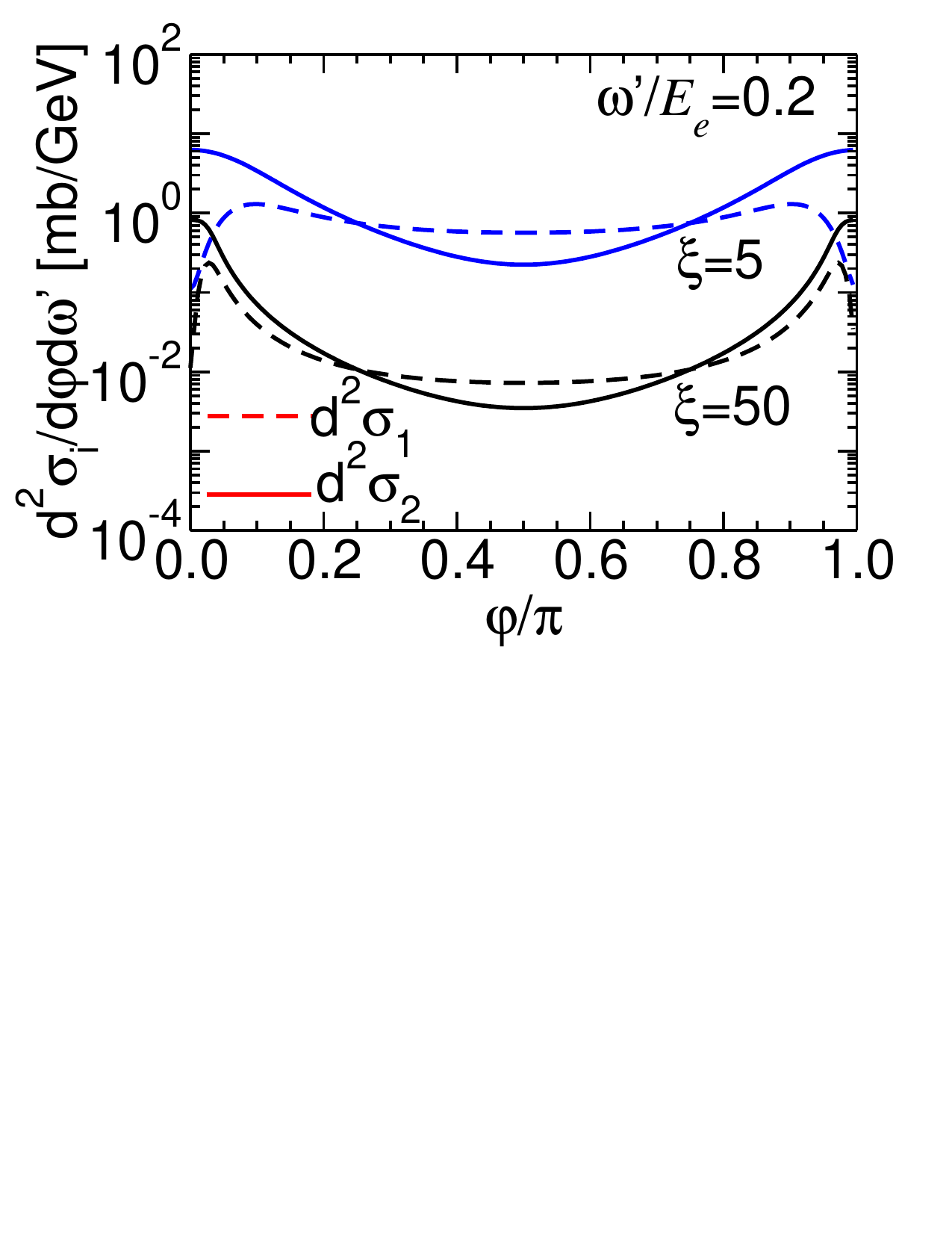}
\includegraphics[width=0.49\columnwidth]{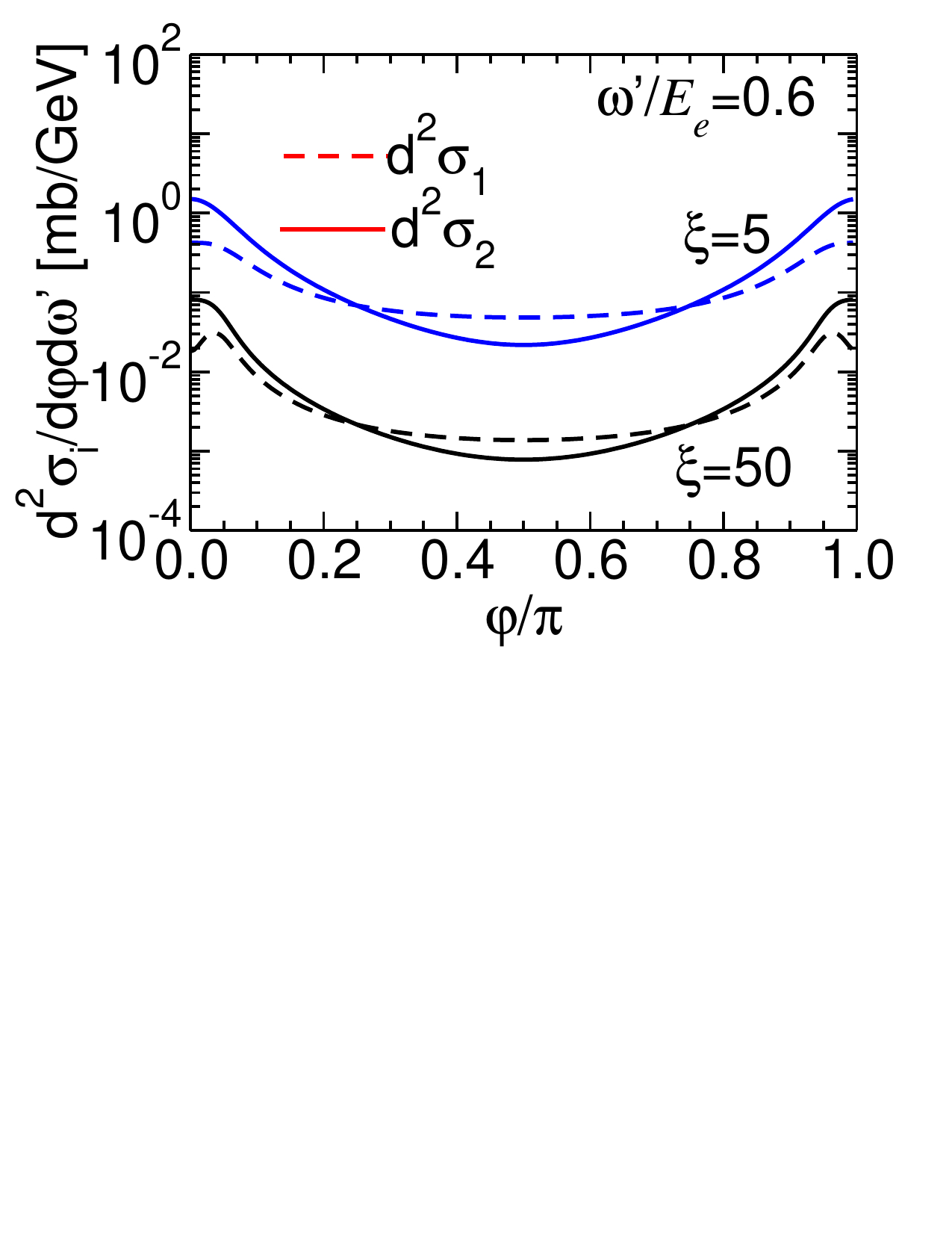}
\vspace*{-3.4cm}  
\caption{\small{(Color online)
    The left and right panels correspond to the azimuthal
    angle distributions of the partial differential cross sections
    $d^2\sigma_i/d\omega'$ for $\omega'/E_e=0.2$ and 0.6, respectively,
    for $\xi=5$ and 50.
    \label{Fig:11}}}
\end{figure}
The partial distributions $d\sigma_{1,2}/d\omega'$ integrated over the azimuthal
angle $\varphi$ for $\xi=5$ and 50 are shown
in the right panel of Fig.~\ref{Fig:10}. One can see
that $d\sigma_2>d\sigma_1$.
However, it is impossible to determine the alignment of
$\gamma'$-polarization only from the predominance of $\sigma_2$,
averaged over the azimuthal angle.

For this purpose, Fig.~\ref{Fig:11} shows the distributions
of partial cross sections as functions of the azimuthal angle
for $\omega'/E_e=0.2$ and 0.6 exhibited in the left and right panels,
respectively.
Note that the structure of $d^2\sigma_i$ confirms the symmetry
property (\ref{symmetry}).

\begin{figure}[h]
\includegraphics[width=0.49\columnwidth]{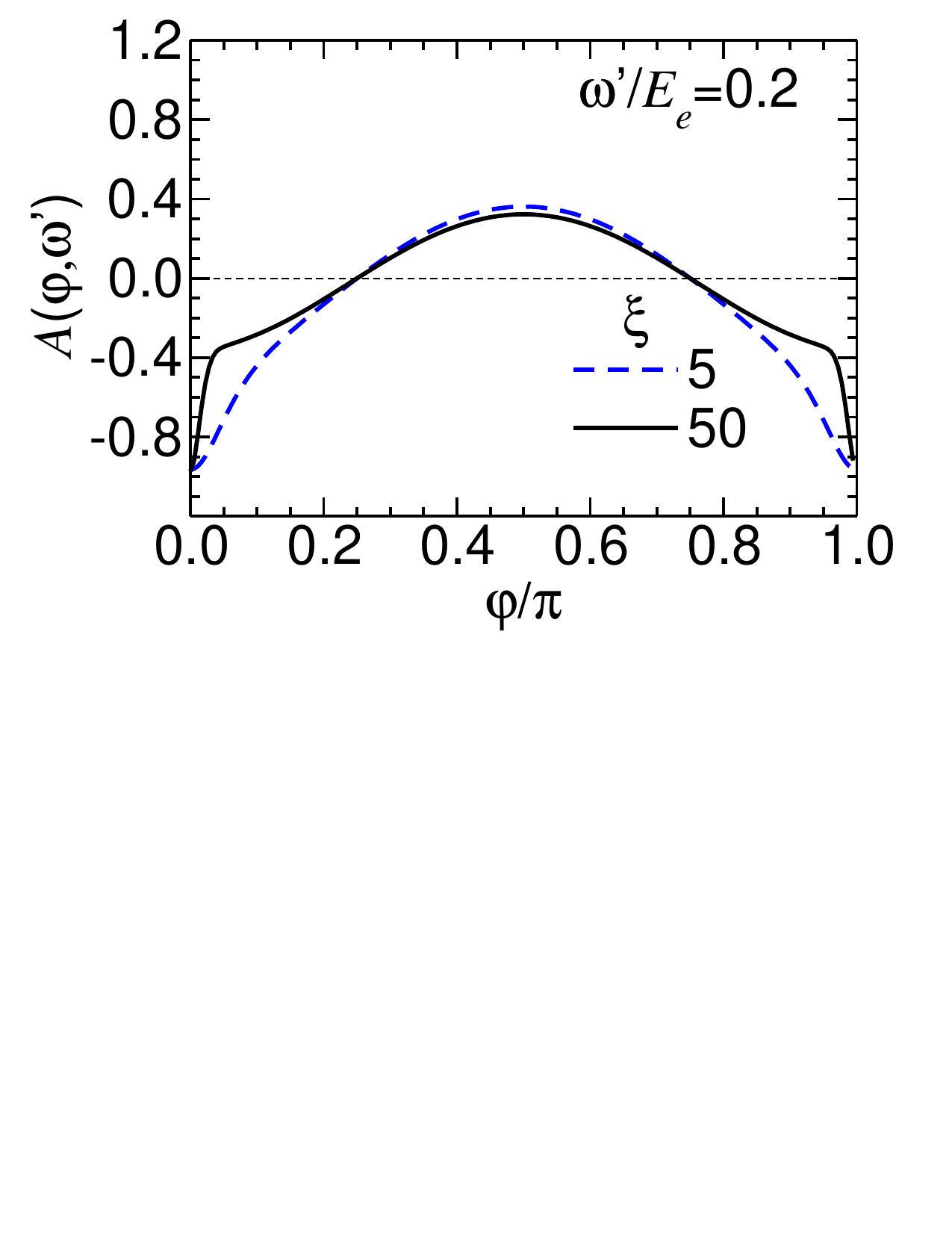}
\includegraphics[width=0.49\columnwidth]{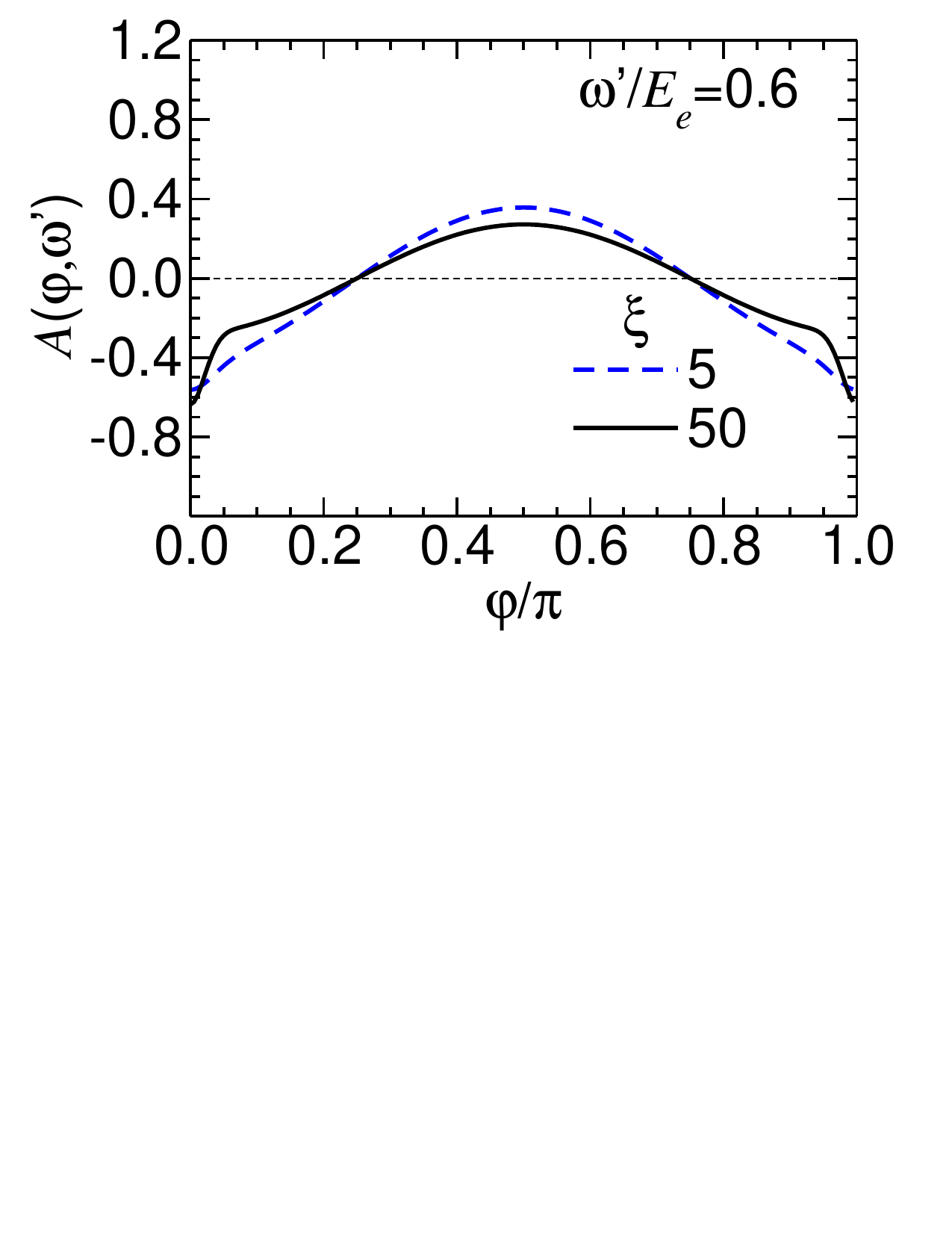}  
 \vspace{-3.4cm} 
 \caption{\small{(Color online)
The asymmetries ${\cal A}(\varphi,\omega')$ as functions of the azimuthal
angle $\varphi$ at $\omega'/E_e=0.2$ and 0.6 exhibited in the left and
right panels, respectively.
\label{Fig:12}}}
\end{figure}

The corresponding asymmetries are shown in Fig.~\ref{Fig:12}. 
In the vicinity of $\varphi=0\,(\pi)$, $d^2\sigma_2\gg d^2\sigma_1$
$({\cal A}\sim -1)$,
which leads to alignment of the $\gamma'$ polarization along
the {laser beam} polarization. The degree of polarization
${\cal P}$ is close to unity. 
These areas provide the main contribution
to the integral cross sections.
On the other hand, in vicinity of $\varphi=\pi/2$, $d^2\sigma_1$ dominates
(${\cal A}>0$).
This again leads to the alignment of $\gamma'$ polarization along the
{laser} beam polarization.    
    
\begin{figure}[th]
 \includegraphics[width=0.48\columnwidth]{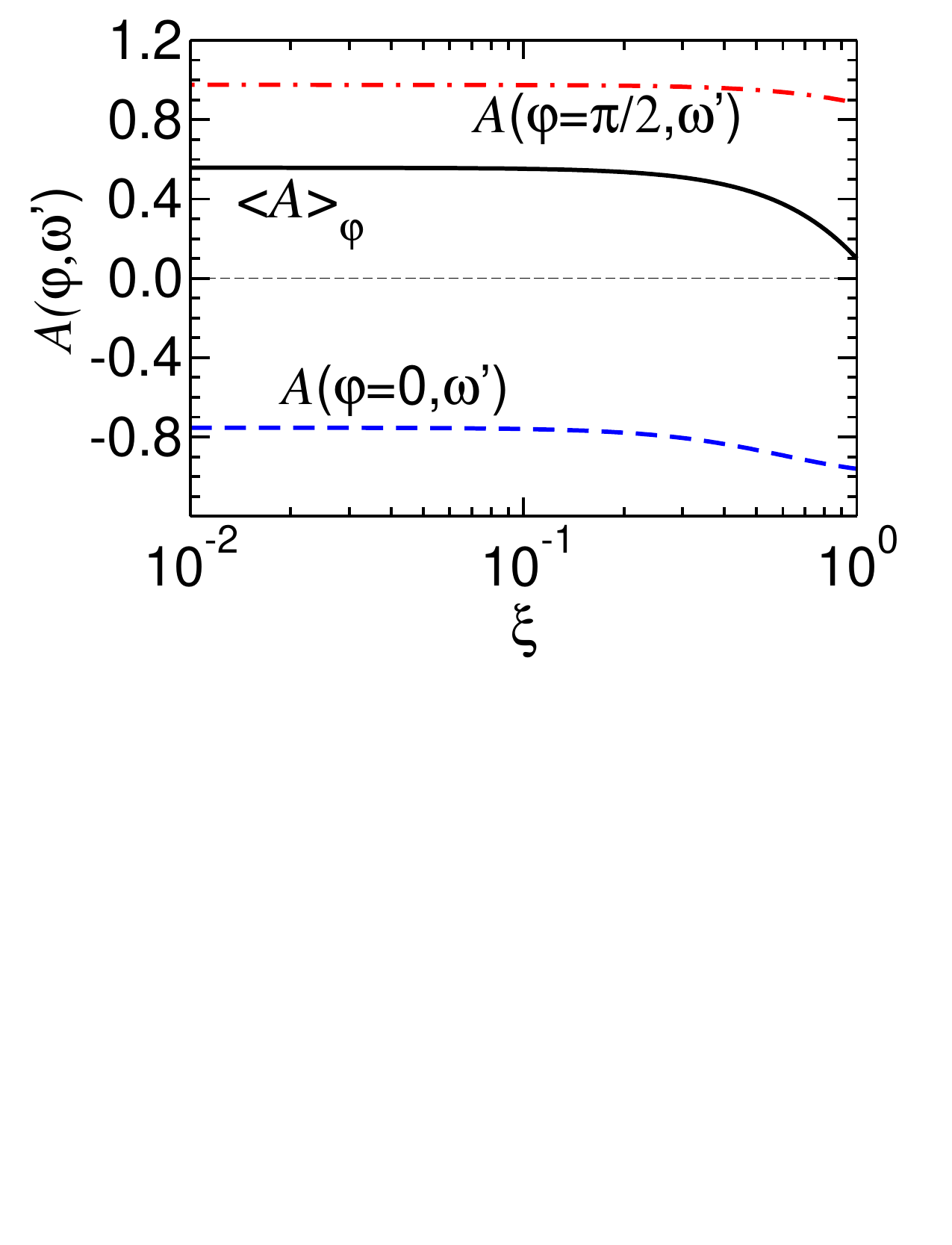}
 \includegraphics[width=0.48\columnwidth]{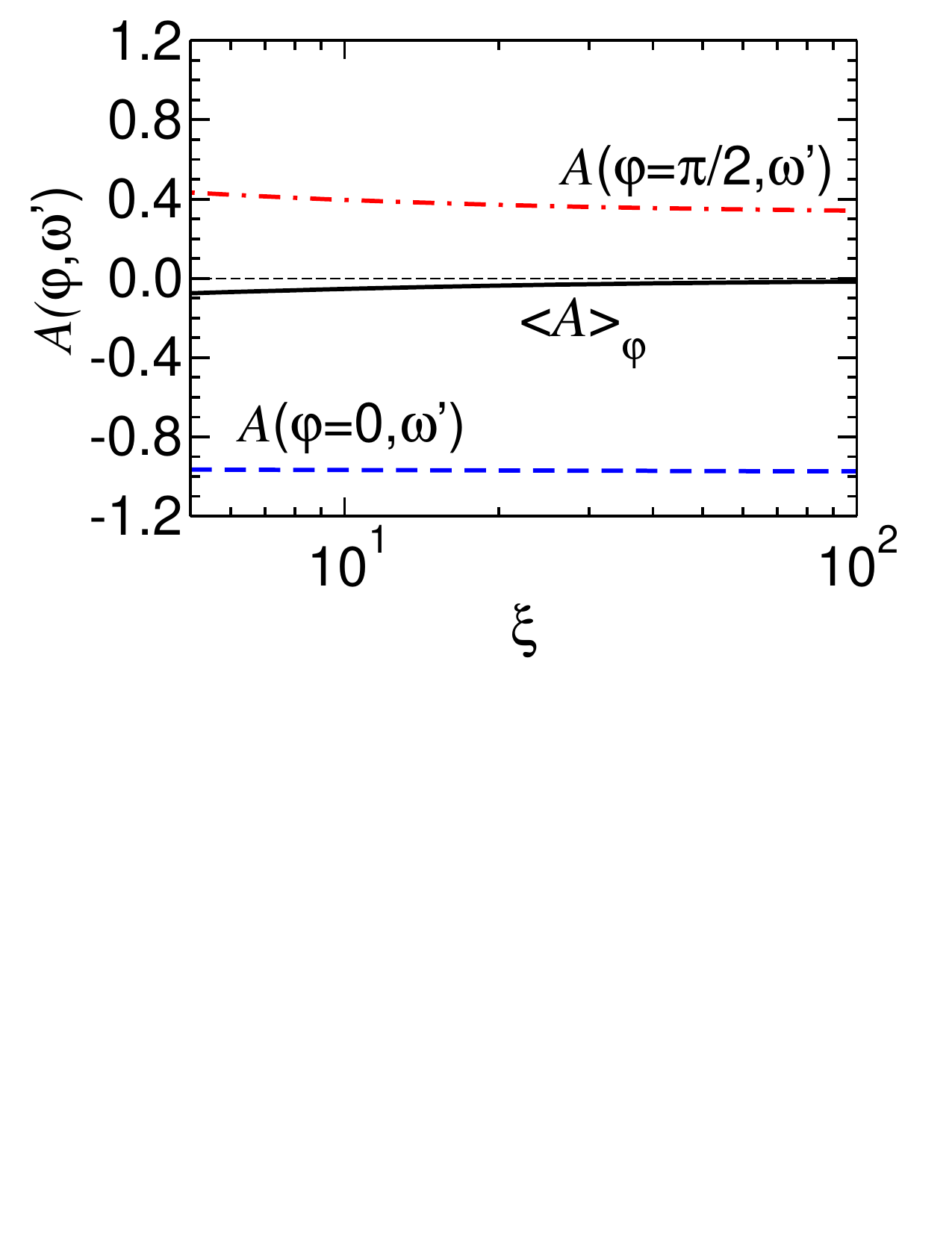} 
\vspace*{-2.8cm}  
\caption{\small{(Color online)    
    The asymmetries ${\cal A}(0,\omega')$, ${\cal A}(\pi/2,\omega')$, and
    the average asymmetries $\langle{\cal A}\rangle_\varphi$ 
    at small and large values of $\xi$, shown in the left and
    right panels, respectively, at
    $\omega'/E_e=0.2$.
 \label{Fig:13}}}
\end{figure}
For convenience, the asymmetries
${\cal A}(0,\omega')$, ${\cal A}(\pi/2,\omega')$, and
$\langle{\cal A}\rangle_\varphi$ 
at $\omega'/E_e=0.2$ in a wide region of $\xi$
are shown in Fig.~\ref{Fig:13}.
The left and right panels correspond to the intervals 
$0.01\le\xi\le 1$ and $5\le\xi\le100$, respectively.
The asymmetry in coplanar geometry ${\cal A}(\varphi=0,\omega')$
is negative with a large absolute value regardless of the field strength.
When $\varphi=\pi/2$, the asymmetry ${\cal A}(\varphi=\pi/2,\omega')$
is positive, while the average asymmetry varies from large to small
values, when $\xi$ varies from small to large values $\xi$.
At $\xi\gg1$, the average asymmetry is close to zero, which is a consequence
of the azimuthal angle distributions exhibited in Fig.~\ref{Fig:12}. 

\section{Summary}

In summary, we have analyzed polarization of the recoil
photon $\gamma'$ in non-linear Compton scattering.
The polarization axes ${\mathbf{e}'_{1,2}}$ are chosen
according to (\ref{07}).
Polarization of $\gamma'$ is described by the asymmetry ${\cal{A}}$
which is equal to the intrinsic spin variable $\xi^{f}_3$ of $\gamma'$.
The initial photon polarization is chosen along the ${\mathbf x}$ axis.

We have analyzed the $\gamma'$ polarization in the LUXE experiment
with the electron energy $E_e=16.5$~GeV and optical laser 
frequency of 1.55 eV in a wide range~of~$\xi$.

The result of our study shows that for small $\xi<1$
and $\omega'/E_e\simeq0.2$
the asymmetry ${\cal{A}}(\varphi,\omega')$ is large and negative
or positive for $\varphi=0 $ or $\varphi=\pi/2$, respectively,
which indicates that the recoil photon is polarized along the $\mathbf{e}'_2$
or $\mathbf{e}'_1$ axes, respectively.
In both cases, $\gamma'$ is polarized
along the {laser} beam polarization and the degree of polarization
${\cal P}$ is close to unity.
The average asymmetry $\langle A \rangle_\varphi$ has a bump-like
behavior with a bump position $\omega'/E_e\approx 0.2$.
The finite pulse  does not
modify qualitatively the result for the monochromatic plane wave background.

Concerning our model for the finite pulse,
 it should be noted that despite the transparency and
validity of our approach, it involves
the calculation of multiple integrals with rapidly oscillating functions,
which requires a large amount of computational resources,
especially for a large number of oscillations $N$ in the pulse
and large $ \xi$.
This causes inconvenience, say, in using our approach as an input
for kinetic codes. Nevertheless, it can be used as a test for
approximate approaches such as the local-constant-field approximation
(LCFA)~\cite{Ritus-79,Piazza2018} or "locally monotonic''
approximation (LMA)~\cite{Heinzl2020};
for LCFA and LMA, see review~\cite{AM_Review} and references therein.
On the other hand, our results for a finite pulse
coincide qualitatively with the results of the monochromatic PW background,
so the latter can be used in qualitative estimations.

At ultra-high laser {beam} intensity with
$\xi\gg1$ the asymmetry ${\cal{A}}(\varphi,\omega')$  is negative (positive)
at $\varphi=0$ ($\pi/2$).
In both cases, $\gamma'$ is polarized
along the {laser beam} polarization.
In case of $\varphi=0\,(\pi)$ the degree of polarization is close to unity.
The average asymmetry is close to zero.

Our calculation was performed for polarization axes in the form
(\ref{077}), (\ref{07}).
A different choice of axes (see for example~\cite{Seipt:2020diz})
would lead to a different dependence of partial cross sections
on the azimuthal angle, which can be the subject of
experimental study of this issue.
However, the main conclusions of~\cite{Seipt:2020diz}
relative to the directions and degree of $\gamma'$ polarization
are close to ours.

In spite of the fact that the partial cross sections  in III~B, III~C and
IV are expressed in a similar form, they are different because
correspond to the different initial conditions: monochromatic background,
background with the finite pulse, and large field intensity with $\xi\gg1$,
respectively. 
The difference appears in a different structure of the basis functions,
which leads to different methods of calculating multidimensional integrals
for cross sections.
We choose such presentation as a transparent
and convenient for those who want to repeat our results.

Finally note that our result assumes the preferred configuration
of experiment on $\ee$ pair production in the two-step electron-laser
interaction with linearly polarized laser beams
(see, for example, Fig.~3 in~\cite{LUXE_exp}).
The first and the second steps are the nlCo and nlBW processes, respectively.
The mutual polarization of these two {laser} beams must be
perpendicular since this configuration leads to enhanced 
yield of $\ee$ pair production~\cite{RitusGroup,TK2020}.

The yield of $\ee$ pairs in electron-laser interaction
based on the nlCo$\otimes$nlBW-type folding model was estimated
at a probabilistic
level in~\cite{TAK2021}. The polarization of recoil photons was not taken
into account.
Accounting for the $\gamma'$ polarization can be
a further development of such studies.

\section{Acknowledgments}

I am grateful to B. K\"ampfer for our fruitful previous collaboration
on studying different topics of strong-field QED
and to O.~V.~Teryaev for discussions of various aspects of spin physics.\\

Data Availability Statement: No Data associated in the manuscript.

\end{document}